# Title

Surviving the frailty of time to event analysis in massive datasets with Generalized Additive Models (and the help of Simon Laplace)

## Authors

Christos Argyropoulos MD, MS, PhD, FASN [1,2,*], Hamza Mir MS [1], Maria Eleni Roumelioti MD [1] Pablo Garcia MD[1]

## Affiliations

[1]Department of Internal Medicine, Division of Nephrology, 1 University of New Mexico MSC 04-2785, Albuquerque, New Mexico, 87131, United States of America

[2]UNM Clinical and Translational Science Center, 900 Camino de Salud, Albuquerque, NM 87131, United States of America

[*] Corresponding author: cargyropoulos@salud.unm.edu

# Abstract

Analyses of time to event datasets have been invariably based on the Cox proportional hazards model (PHM). Reformulations of the PHM as a Poisson Generalized Additive Model (GAM) or as a Generalized Linear Mixed Model (GLMM) have been proposed in the literature, aiming to increase the flexibility of the PHM and allow its use in situations in which complex spatiotemporal relationships have to be taken into account when modeling survival. In this report, we provide a unified framework for considering these previous attempts and consider the implementation in software for GAM and GLMM in the R programming language. The connection between GAM/GLMM and the PHM is leveraged to provide computationally efficient implementations for a subclass of survival models that incorporate individual random effects ("frailty models"). Frailty models provide a unified method to address repeated events, correlated outcomes and also time varying visitation schedules when analyzing Electronic Health Record data. However the current implementation of frailty models in software facilities for the Cox model does not scale because of long computation times; conversely the direct implementation of individual random effects in GAM/GLMM software does not scale well with memory usage. We propose a two stage method for survival models with frailty based on the Laplace approximation. Using a D-optimal experimental design to simulate the performance of the proposed method across simulated datasets we illustrate that the proposed method can circumvent the limitations of existing implementations, opening up the possibility to model datasets of hundred of thousands to million individuals using high end workstations from within R.

# 1. Introduction

More than fifty years since its introduction by Sir Cox(1), the proportional hazards model remain the dominant approach to analyzing time to event ("survival") data, as it leads to easily interpretable estimates of relative effects, e.g. how more or less likely is the presence of a given feature to be associated with a particular observable outcome. The major statistical innovation of the model, i.e. the introduction of the partial likelihood concept(2), which was also a major technical innovation to speed up calculations was the description of estimation procedures for relative effects that bypassed the more numerically involved estimation of the time varying risk of the referent group ("baseline hazard") in a given analysis. Equivalent formulations based on Generalized Linear Models(3, 4) were described soon after the publication of the proportional hazards model (5–10) and were in fact used to carry out survival analysis with the proportional hazards model before the introduction of dedicated fitting facilities in statistical software packages (11–15). However, these approaches did not gain much traction, likely because they were felt to be heuristic (p 358, (16)) approaches lacking rigor, and computationally demanding for the computers available the 1980s (10).

Nevertheless, there is considerable merit in further exploring the relationship between survival analysis and GLMs: notably, modeling complex non-linear functional, temporal, or spatio-temporal relationships will require the explicit modeling of the hazard function that the proportional hazards model deliberately chooses to leave unspecified. Motivated by such considerations, we (17) derived from first principles a new perspective in the GLM-survival connection by relating methods for numerical integration of the hazard function to exposures in contingency tables. This new perspective allowed us to not only provide a bound to the approximation error of survival models via GLMs, but clearly demarcated the computational resources needed to leverage the connection between survival analysis and penalized regression methods for GLMs i.e. Generalized Additive Models (GAMs, (18)) as implemented in the R package *mgcv*(19).

In this report we revisit the GAM survival connection in the era of big electronic health record (EHR) data. Such data are too big, grow at high velocities and are by definition partial as a large (possibly the largest) component of the data managed by healthcare organizations lacks structured representation despite initiatives to organize the chaos(20, 21). As a result of the partial representation of clinical information, *any* model that one would hope to build, will exhibit substantial heterogeneity in the occurrence of the outcome of interest when individuals with the same levels of observed covariates (e.g. gender, race, socioeconomic status) are noted to have vastly different times until they experience an event of interest(22–24). Such heterogeneity is also observed in the timings between recurrent clinical events *within* the same individual and is known as *frailty* (25). Despite the clear methodological pathway for frailty models, such techniques are extremely demanding computationally and their performance for giga-datasets has not been considered for either extensions of the proportional hazards model or the GAM interpretations. We thus decided to extend our previous explorations of random effects modeling with Generalized Linear (Mixed) Models(26) in EHR to survival analysis. We illustrate through simulations that current gold standard implementations of frailty models in R (packages *survival* (27) and *coxme*(28)) but particularly the direct implementation of subject level frailties as random effects in *mgcv* will not scale with the size of the dataset. Motivated by these observations, we propose a method that decouples the estimation of subject level frailty factors from the flexible estimation of survival, and allows the scalable estimation of frailty models with hundreds of thousands random effects within R using readily available workstations.

## 2. Theory and Methods

By way of introduction we present a complete derivation of the Poisson connection to survival analysis without relying at all on counting processes or martingales in section 2.1. Of historical interest, versions of their derivation may be found in some old textbooks (e.g. pages 74-76 and 358-363 in (16), but not in the corresponding sections in pages 351-354 of the 2$^{nd}$ edition with both editions labeling the key references by Holford (8) and Breslow(6) heuristic). Using standard calculus (mean value theorem for integrals) we show that these derivations are in fact quite rigorous and provide exact GLM likelihoods for piecewise continuous (not just piecewise constant) hazards: piecewise constancy is required *only* to provide estimates and predictions involving the survival function. We conclude the section by showing how the Gaussian integration rules we presented previously allow the GLM reformulation of the survival analysis to be exact for a wide range of continuous hazards, while controlling the computational resources required. In section 2.2. we discuss the estimation process of the Poisson survival model GLMM (*glmmTMB*) and GAM (*mgcv*) software in the R language and potential limitations for the direct implementation of frailty survival models and outline the design of pilot experiments to map the scalability of the reference packages for fitting frailty models in R, *coxme* and *coxph* and the *mgcv* implementations. In section 2.3 we propose a general backfitting algorithm that can dramatically cut down on the time and memory resources required to fit extremely large frailty models through a rational combination of distinct software components for the survival and frailty estimation part. Section 2.4 presents the design of a D-optimal response surface simulation experiment in which we mapped the scalability (memory resources and execution timings) of the proposed backfitting approach and the direct implementation of frailty survival models using *glmmTMB*. This experiment is used to select an optimal combination of software components for frailty modeling by mapping resource use and the bias of the estimators generated by the various frailty models for very large datasets with variable number of covariates.

## 2.1. Jumping hazards without semi-parametrics using quadratures

In its most common application survival analysis deals with collections of observation times of individuals, the times these individuals came to observation (entry points), and censoring indicators that tell us whether an individual experienced the event of interest (indicator assumes the value of one) or not (indicator equal to zero) at the recorded observation time. Under the assumption of a *non-informative censoring*, the likelihood of the collection of observation times ($T_i$), entry points ($E_i$) and censoring ($\delta_i$) indicators is given by

$$\prod_{i=1}^{N} \frac{f(T_i)^{\delta_i} \times S(T_i)^{1-\delta_i}}{S(E_i)} = \prod_{i=1}^{N} h(T_i)^{\delta_i} \times \exp\left(-\int_{E_i}^{T_i} h(t)\, dt\right) \quad \text{Eq 1}$$

with $f(T_i)$ the probability density function, $S(E_i) = \int_0^{T_i} f(t)dt$ the survival function and $h(t)$ the hazard function. In the proportional hazards model these parameters quantify the effects of *risk factors* which are assumed to modify an unknown baseline hazard function $h_0(t)$ in a multiplicative manner, i.e.

$$h_i(t) = h_0(t) \times \exp(x_{1,i}\beta_1 + x_{2,i}\beta_2 + \cdots + x_{p,i}\beta_p) = h_0(t) e^{X_i \beta} \quad \text{Eq 2}$$

One may assume that without an explicit specification of the baseline hazard $h_0(t)$, the integrals in Eq 1 could not evaluated and thus $h_0(t)$ could not be estimated in a data driven fashion, but there is a way. Start by ordering all entry and observation times in the dataset from smallest to largest, and denote the $K$ distinct such times $t_k, 1 \leq k \leq K$. This ordering implies that the integral for the ith individual can be written as the sum of integrals over all the entry and observation times in the dataset that lie between $E_i$ and $T_i$, by jumping from $t_k = E_i$ to $t_{k+1}$, the 1st time (either entry or observation) that follows $E_i$ in the dataset, then to $t_{k+2}$, i.e. the 2nd time following $E_i$ until one reaches the end of the observation for that individual at time, having encountered $\kappa_i$ distinct times $t_{k+\kappa_i} = T_i$.

$$\int_{E_i}^{T_i} h_i(t)dt = \sum_{j=0}^{\kappa_i - 1} \int_{t_{k+j}}^{t_{k+j+1}} h_i(t)dt = e^{X_i \beta} \sum_{j=0}^{\kappa_i - 1} \int_{t_{k+j}}^{t_{k+j+1}} h_0(t)dt \quad \text{Eq 3}$$

At this point, one invokes the mean value theorem of integrals to rewrite Eq 3, as

$$\int_{t_{k+j}}^{t_{k+j+1}} h_0(t)dt = h_0(\xi_{k+j}) \times \underbrace{(t_{k+j+1} - t_{k+j})}_{\Delta t_{k+j}} = \overline{h_0}(t_{k+j} \to t_{k+j+1})\Delta t_{k+j}, \ t_{k+j} \leq \xi_{k+j} \leq t_{k+j+1} \ \text{Eq 4}$$

The mean value theorem in this case, allows us to replace the (unknown) value of the integral of the hazard, i.e. the cumulative hazard, by the product of the length of the interval of the integration ($\Delta t_{k+j}$) and either the average value of the hazard in that interval $\overline{h_0}(t_{k+j} \to t_{k+j+1})$ or the value of the hazard at an (unknown) point inside the interval $h_0(\xi_{k+j})$. The construction implied by Eq 3 is exact but is not unique: one could choose to partition time not at every single observation time but only at those in which an event occurred. In fact, let's do this for the case of right censored data, i.e. when all individuals enter observation at time zero. Doing so replaces the problem of integrating an unknown function (the baseline hazard), with the problem of evaluating an unknown function, $h_0(\cdot)$, over an irregular grid of unknown points ($\xi_{k+j}$). Denote $h_{0,k+j} = h_0(\xi_{k+j})$ the values of the hazard that verify Eq 4 in each of the intervals $\{(t_k, t_{k+1}]\}_{k=0}^{K-1}$ with $t_0 = 0$. Introduce the indicators $I_{i,k} = 1$, $\delta_{i,k} = 1$ if the ith individual was under observation and experienced the event of interest in the $(t_k, t_{k+1}]$ respectively, and the corresponding risk and death sets $R_k, D_k$. Then **Eq 1** may be reformulated as:

$$\prod_{i=1}^{N} \prod_{k=1}^{K} \left(h_{0,k} e^{X_i \beta} \Delta t_k\right)^{I_{i,k}\delta_{i,k}} \exp\left(-I_{i,k} e^{X_i \beta} h_{0,k} \Delta t_k\right) \ \text{Eq 5}$$

which after absorbing the indicators $I_{i,k}$ as tests of membership in the risk sets may be rearranged to:

$$\prod_{k=1}^{K} \prod_{i \in R_k} \left(h_{0,k} e^{X_i \beta} \Delta t_k\right)^{\delta_{j,k}} \exp\left(-e^{X_i \beta} h_{0,k} \Delta t_k\right) \ \text{Eq 6}$$

To make the connection with the Cox estimation we will examine the logarithm of Eq 6, which is within an additive constant the log-likelihood of Eq 1:

$$LL(\beta, h_{0,1}, h_{0,2}, \ldots, h_{0,k}) = \sum_{k=1}^{K} \left[d_k \log h_{0,k} + \sum_{i \in D_k} X_i \beta - h_{0,k} \sum_{i \in R_k} e^{X_i \beta} \Delta t_k\right] \ \text{Eq 7}$$

In this expression $d_k = \sum_{i=1}^{N} \delta_{i,k}$ is the number of deaths between $t_k$ and $t_{k+1}$, $R_k$ the risk set, i.e. the individuals who were under observation and thus at risk of dying in that interval and $D_k$, the death set, i.e. the individuals who actually died in that interval. Using elementary differential calculus, Holford(7, 8) and Breslow(6) showed five decades ago, that optimizing the Poisson construction in Eq 5 re-expressed as Eq 7 leads to the 1) same estimates as the proportional hazards model for the (log-)relative risks (the $\boldsymbol{\beta}$) 2) the Breslow estimator for the baseline hazard and 3) the Breslow method for handling tied observations. For the sake of completeness, we outline the steps in the derivation:

1) first differentiate Eq 7 with respect to $h_{0,k}, 1 \leq k \leq K$ to arrive at $h_{0,k} = \frac{d_k}{\sum_{i \in R_k} e^{X_i \beta} \Delta t_k}$ (this becomes the Breslow estimator upon convergence)

2) Then substitute back to the expression of the score equation $\boldsymbol{U}(\boldsymbol{\beta})$ (the vector of first derivatives of the log-likelihood in Eq 7) leading to: $\boldsymbol{U}(\boldsymbol{\beta}) = \sum_{k=1}^{K} \left\{ \sum_{i \in D_k} \boldsymbol{X}_i - d_k \frac{\sum_{i \in R_k} X_i e^{X_i \beta}}{\sum_{i \in R_k} e^{X_i \beta}} \right\}$ after cancellation of the length of the intervals from the numerator and denominator of the fraction. This is the score equation that is used to estimate parameters in the Cox model in the absence of ties ($d_k = 1, \forall k$) and the Breslow estimator for the proportional hazards model in the presence of ties. If one denotes the estimates of $\boldsymbol{\beta}$ as $\widehat{\boldsymbol{\beta}}$, then one can estimate $h_{0,k}$ as $\widehat{h}_{0,k} = \frac{d_k}{\sum_{i \in R_k} e^{X_i \widehat{\beta}} \Delta t_k}$. Standard errors for $\widehat{\boldsymbol{\beta}}$ and may $\widehat{h}_{0,k}$ be obtained from the observed information (the second derivative of the log likelihood).

A few comments are warranted here: firstly, that the derivation does not make any assumption about the hazard function (it only requires that the baseline hazard be at least piecewise continuous) and thus the results obtained are quite general. Secondly, the relative risk estimates are obtained through standard maximum likelihood but are identical to what the partial likelihood of the Cox model provides, illustrating the efficiency of the later. A similar result was obtained by considering marginal likelihoods for rank statistics (29). Finally, the survivor function (or functions that depend on it) cannot be determined without further assumptions about the baseline hazard that will allow one to map $\widehat{h}_{0,k}$ to $\xi_k$ and thus "draw" the survival curve at different time points. The assumption made in semiparametric/counting process expositions is that the baseline hazard stays constant between consecutive event times. Under this assumption the mean value theorem of Eq 4 is satisfied for any $\xi_k$ in the interval $(t_k, t_{k+1}]$ leading to the familiar pattern of stepwise declining survival function (and the not so familiar pattern of the jumping hazards).

Eq 5 and Eq 6 also provide the justification for fitting Cox models using software for Poisson GLM using standard maximum likelihood. The key observation is that the inner product in Eq 6 is the likelihood of a Poisson GLM in the canonical (log) link with intercept $\log h_{0,k}$ and offset $\log \Delta t_k$. Now define 1) an outcome variable $y_{i,k} = I_{i,k}\delta_{i,k}$ that assumes the value of zero when either the ith individual has experienced no events under observation, or if they were not under observation in the kth time interval and one if they experienced an event in the kth interval 2) a vector of parameters for the log hazards $\lambda = (\log h_{0,1}, \log h_{0,1}, \ldots, \log h_{0,K})$ which correspond to $h_0(\xi_{k+j})$ or $\overline{h_0}(t_{k+j} \to t_{k+j+1})$, depending on the two equivalent mathematical forms of the mean value theorem one would like to adopt. Using this notation, we can formally rewrite Eq 5 as

$$\prod_{i=1}^{N}\prod_{k=1}^{k_i}\left(e^{X_i\beta + \Lambda_i \lambda + \log \Delta t_k}\right)^{y_{i,k}} \exp\left(-e^{X_i\beta + \Lambda_i \lambda + \log \Delta t_k}\right) \text{ Eq 8}$$

with $k_i$ the last interval in which the ith individual is at risk and $\Lambda_i = [I_{k_i} \quad 0_{k_i \times (K-k_i)}]$ a block $k_i \times K$ matrix, where $I_{k_i}$ is the identity matrix in $k_i$ dimensions and $0_{k_i \times (K-k_i)}$ is a matrix of zeros. Eq 8 may be recognized as a regression model for $\sum_{i=1}^{N} k_i$ Poisson variables. To construct the dataset for this regression one has to split the follow time for each individual at each event time preceding and up to final observation time for that individual, include artificial Poisson zero counts for all the intervals at which the individual was under observation without experiencing an event, and expand the design matrix $X_i$ with $\Lambda_i$. The coefficients for these additional columns provide maximum likelihood estimates for the baseline log hazards based on the individuals who were at risk during those intervals(10, 15, 30).

In passing note, this trick of splitting the observation time to distinct intervals is how one introduces time varying covariates in the Cox model and hence this extension is readily accommodated in the GLM framework.

The GLM reformulation of a proportional hazards model as derived suffers from two theoretical and one computational shortcoming: firstly, the construction is stochastic (or dataset specific) as it depends on the actual event times observed in the dataset to reconstruct the survivor function. Secondly, forcing the hazard to "jump" is not a very realistic assumption for many (or most) biomedical phenomena, in which the baseline hazard varies smoothly, rather than abruptly. The computational limitation follows from expanding the dataset with additional rows (the pseudo-observations) and columns; at some point one runs out of memory to store and manipulate the $\left(\sum_{i=1}^{N} k_i\right) \times (p + K)$ matrices needed to fit the GLM. In addition, not accounting for the block structure and the sparsity of the design matrix $\begin{bmatrix} X_1 & \Lambda_1 \\ \vdots & \vdots \\ X_N & \Lambda_N \end{bmatrix}$ will lead to wasteful computations even if one had large amounts of computer memory at their disposal. We can simultaneously address theoretical and computation issues by NOT using the mean value theorem to handle the integrals in Eq 1. Rather we will use a Gaussian quadrature rule to write the integral as a weighted sum of the values of the baseline hazard over a set of nodes $n_i$ inside the domain of integration and a remainder term:

$$\int_{E_i}^{T_i} h_i(t)dt = \sum_{j=1}^{n_i} w_{i,j} h_i(t_{i,j}) + R_i = \text{Eq 9}$$

Gaussian quadrature techniques (see(31) pages 179-193) allow for the *exact* ($R_i = 0$) integration of polynomials of very high degree given a sufficient number of nodes, e.g. the Gauss Lobatto rule with $n_i$ nodes is exact for polynomial hazards up to degree of $2n_i - 3$, while providing a very accurate result for functions that are well approximated by polynomials. Switching integration techniques allows a much wider class of hazards to be handled *exactly* by the GLM reformulation of the survival analysis and cuts down dramatically the approximation error of forcing functions that are not jumping to jump. In fact, it is not strictly necessary to use a variable number of nodes for e.g. individuals with long observation times, and one may settle for the same value ($n$) for all individuals. Ignoring the remainder term, which will be zero for a large class of hazard functions, the likelihood in Eq 1 retains its Poisson like form the Gauss Lobatto rule:

$$\prod_{i=1}^{N} \prod_{j=1}^{n} \left(e^{X_i\beta+\log h(t_{i,j})+\log w_{i,j}}\right)^{y_{i,j}} \times \exp\left(-e^{e^{X_i\beta+\log h(t_{i,j})+\log w_{i,j}}}\right) \text{ Eq 10}$$

Standard Poisson GLM software can still be used to fit Eq 10 by expanding the dataset with pseudo-observations $y_{i,j} = 0$ at the times corresponding to the internal nodes of the integration rule and $d_{i,n} = \delta_i$ using the logarithm of the quadrature weights ($w_{i,j}$) as offsets. However instead of expanding $\lambda(t_{i,j}) = \log h(t_{i,j})$ as a piecewise constant function, we use an additive linear functional for $\log h(t_{i,j})$ such as a low degree polynomial or a spline to reduce the demand for computational resources. For example, a cubic polynomial would add three columns to the design matrix $X_i$, while a fixed number of nodes scales predictably the rows of matrices used in computations to an integer multiple of the number of individuals. At the conclusion of the model fit, one ends with maximum likelihood estimates (and standard errors) for relative risks, the baseline hazard/survival function and/or any predicted quantity of interest. The estimation process is guaranteed to be exact for a wider class of situations than the standard Kaplan Meier curve (or the semi-parametric proportional hazards model) in a manner that is controlled by the analyst: when in doubt, just increase the number of nodes in the Gauss Lobatto quadrature.

### 2.2. Flexible survival analysis using Generalized Linear Mixed and Additive Models

While survival data can be analyzed via conventional Poisson GLMs it is often advantageous to estimate the baseline hazard via penalized regression models that for example express a belief that the hazard be a smooth rather than a wiggly, oscillating function. To do so, we will penalize some of the parameters ($b_R$) in the linear expansion of the log-hazard, $\log h(t_{i,j}) = X_{i,F}b_F + X_{i,R}b_R$ with $X_{i,F}$ and $X_{i,R}$ design matrices that are evaluated at the time points $t_{i,j}$, and the penalty assuming a multivariate Gaussian form. In this case the penalized likelihood is within a proportional constant the joint likelihood of a GLMM for some functional specific precision matrix $S$ and penalty $\lambda$:

$$\prod_{i=1}^{N}\prod_{j=1}^{n}\left(e^{X_i\beta+X_{i,F}b_F+X_{i,R}b_R+\log w_{i,j}}\right)^{y_{i,j}} \times \exp\left(-e^{e^{X_i\beta+X_{i,F}b_F+X_{i,R}b_R+\log w_{i,j}}}\right) \times \exp\left(-\frac{\lambda}{2}b_R^T S b_R\right) \text{Eq 11}$$

Using simulations and real world clinical trial datasets we showed in (17) that moderate discretization, e.g. expanding its lifetime to 10-20 pseudoobservations and application of the package *mgcv* to fit Eq 11 as a GAM via the Laplace Approximation (see 2.3 for some of the computational details), allows one to estimate proportional hazards effects and the survivor function in an unbiased manner and with acceptable coverage. The results are virtually indistinguishable from those obtained by the proportional hazards model and the Kaplan Meier estimator. However, Eq 11 defines the joint likelihood of a GLMM and this view allows estimation to be carried out with any GLMM software package. In addition to the original function *gam* in *mgcv*, further options include the big-dataset oriented function *bam*(32) in the same package, and the packages *glmmTMB*(33) and *gamlss*(34). Table 1 summarizes an extensive, but not exhaustive list survival analysis scenarios that are supported by the *mgcv* and *glmmTMB* model statement interface after one has expanded the dataset as in the previous section.

**Table 1** *mgcv/glmmTMB model statements for survival analysis scenarios*

| Scenario | Model statement |
|---|---|
| Unadjusted proportional hazards model for treatment | `y_ij ~ s(t) + treat + offset(log(w_ij))` |
| Treatment and linear age effect | `y_ij ~ s(t) + treat + age + offset(log(w_ij))` |
| Treatment and flexible/non-linear age effect | `y_ij ~ s(t) + treat + s(age) + offset(log(w_ij))` |
| Unadjusted proportional hazard analysis stratified by center | `y_ij ~ s(t, by=clinic) + treat + clinic + offset(log(w_ij))` |
| Multiple time scales and treatment effect | `y_ij ~ s(t) + s(period) + treat + offset(log(w_ij))` |
| Age-period model with smoothing thin plate spline and treatment effect | `y_ij ~ s(t) + s(period, upd_age ,bs="ts") + treat + offset(log(w_ij))` |
| Time-varying treatment effect | `y_ij ~ s(t, by=treat)+ treat + offset(log(w_ij))` |
| Covariate by treatment interaction and overall treatment effect | `y_ij ~ s(t) + s(age , by=treat) + treat + offset(log(w_ij))` |
| Treatment effect in the presence of frailty (modeled as random effect) | `y_ij ~ s(t) + s(id , bs="re") + treat + offset(log(w_ij))` |
| Time varying treatment effect with stratified baseline hazard, flexible age effect and individual frailty | `y_ij ~ s(t, by=treat) + treat + s(id , bs="re") + s(age) + s(t, by=clinic) + clinic + offset(log(w_ij))` |

Hypothetical clinical dataset, that has been expanded to include Poisson pseudo-observations `y_ij` with observation time (`t`) and Gauss Lobatto weights `w_ij`. The dataset includes covariates of `age` at baseline, age updated at the time of each observation (`upd_age`), an intervention assignment (`treat`), calendar time (`period`), the clinical site (`clinic`) that study participants were enrolled at and individual frailty terms (`id`). Individuals are not assumed to all enter at the same time (delayed entry scenarios). All scenarios require the use of the Poisson family under the canonical (log) link to be equivalent to a proportional hazards model. Similar but not identical model statements are supported by the package *gamlss*, but are not considered here further.

When using multiple (e.g. $N_s$ smoothing effects, the likelihood in Eq 11 is augmented by the quadratic penalties for each smooth term; for notational simplicity we introduce the block diagonal matrix $S_\lambda = \sum_{r=1}^{N_s} \lambda_r diag(\mathbf{0}_{R_1}, \ldots, \mathbf{0}_{R_{r-1}}, S_r, \mathbf{0}_{R_{r+1}}, \ldots, \mathbf{0}_{R_{N_s}})$ [1] to represent the penalties for all smoothers, including that of the log-hazard, other predictors (except individual random effects) or interactions of log-hazard with said predictors. We then augment the fixed effects parameter vector to include the parametric components $\boldsymbol{\beta} = (\boldsymbol{\beta}, \boldsymbol{b}_{F,1}, \ldots, \boldsymbol{\beta}\boldsymbol{b}_{F,j})$ with augmented design matrix $X = \begin{bmatrix} X_1 & X_{1,F,1} & \ldots & X_{i,F,N_s} \\ \vdots & \vdots & \vdots & \vdots \\ X_N & X_{N,F,1} & \ldots & X_{N,F,N_s} \end{bmatrix}$

(with $X_{i,F,k}$ the fixed design matrix of the kth smoother for the ith individual) and introduce the parameter $\boldsymbol{\gamma} = (\boldsymbol{b}_{R,1}, \ldots, \boldsymbol{\beta}_{R,j})$ with the corresponding smoother random effects design matrix $\boldsymbol{\Psi} = \begin{bmatrix} X_{1,R,1} & \ldots & X_{1,R,N_s} \\ \vdots & \vdots & \vdots \\ X_{N,R,1} & \ldots & X_{N,R,N_s} \end{bmatrix}$. Individual (or more general unit level) level random effects (e.g. frailty, random intercept and slope models) are introduced into the regression structure via specifying a unit random effects design matrix $\mathbf{Z}$[2], a vector of random effects $\mathbf{b}$ that is distributed according to the MultiVariate Normal (MVN) distribution with a mean of zero and an associated, parameterized covariance matrix $\mathbf{V_\theta}$. Setting $\mathbf{GL} = (\mathbf{GL}_1, \dots, \mathbf{GL}_N)$ the stacked vector of the Gauss-Lobatto weights and $\mathbf{Y} = (\mathbf{y}_1, \dots, \mathbf{y}_N)$ the stacked vector of individual, Gauss Lobatto expanded responses, the mixed model reformulation of Eq 11 is given by:

$$\begin{aligned} g(\mu) = \log \mu &= \mathbf{X}\boldsymbol{\beta} + \boldsymbol{\Psi}\boldsymbol{\gamma} + \mathbf{Z}\mathbf{b} + \mathbf{GL}, \quad \boldsymbol{\gamma} \sim MVN(\mathbf{0}, \mathbf{S}_\lambda^-), \quad \mathbf{b} \sim MVN(\mathbf{0}, \mathbf{V_\theta}) \\ (Y, \boldsymbol{\gamma}, \mathbf{b} | X, \boldsymbol{\Psi}, \mathbf{Z}, \boldsymbol{\beta}, \mathbf{GL}, \lambda, \theta) &= \prod_{i=1}^{N} \prod_{j=1}^{n} e^{y_{i,j}\mu_{i,j}} \times \exp\left(-e^{e^{\mu_{i,j}}}\right) \times \exp\left(-\frac{\boldsymbol{\gamma}^T \mathbf{S}_\lambda \boldsymbol{\gamma}}{2} - \frac{\mathbf{b}^T \mathbf{V}_u^- \mathbf{b}}{2}\right) \end{aligned} \quad \text{Eq 12}$$

---

[1] In this expression, $R_k$ is the dimensionality of the random effects associated with the kth smooth term, and $\mathbf{0}_{R_k}$ is the square matrix of dimensions $R_k \times R_k$, with all elements equal to zero.

In anticipation of the material of the next section, we find it advantageous to give an equivalent representation that folds the individual (unit level) random effects into the smooth effects through the reparameterization: $\boldsymbol{\Psi}' = [\boldsymbol{\Psi} \quad \mathbf{Z}], \boldsymbol{\gamma}' = (\boldsymbol{\gamma}, \mathbf{b}), \mathbf{S}_{\lambda,\theta} = diag(\mathbf{S}_\lambda, \mathbf{V}_\theta^-), \mathbf{S}_{\lambda,\theta}^- \mathbf{S}_{\lambda,\theta} = \mathbf{I}$:

$$\begin{aligned} g(\mu) = \log \mu &= \mathbf{X}\boldsymbol{\beta} + \boldsymbol{\Psi}'\boldsymbol{\gamma}' + \mathbf{GL}, \quad \boldsymbol{\gamma}' \sim MVN(\mathbf{0}, \mathbf{S}_{\lambda,\theta}^-) \\ (Y, \boldsymbol{\gamma}' | X, \boldsymbol{\Psi}', \boldsymbol{\beta}, \mathbf{GL}, \lambda, \theta) &= \prod_{i=1}^{N} \prod_{j=1}^{n} e^{y_{i,j}\mu_{i,j}} \times \exp\left(-e^{e^{\mu_{i,j}}}\right) \times \exp\left(-\frac{\boldsymbol{\gamma}'^T \mathbf{S}_{\lambda,\theta} \boldsymbol{\gamma}'}{2}\right) \end{aligned} \quad \text{Eq 13}$$

The last two equations form the numerical basis of estimation for all scenarios Table 1 in existing software (Eq 13) and in our proposal for scalable estimation of very large frailty survival models (Eq 12).

### 2.3. Mixing software components for scalable inference of frailty survival models

To estimate the parameters $\boldsymbol{\beta}, \boldsymbol{\gamma}', \lambda, \theta$ in Eq 12, one must find the value of these parameters that maximize the multivariate integral of the joint likelihood over the random effects parameters ($\boldsymbol{\gamma}'$, Maximum Likelihood, ML estimation) or over the fixed and random effects parameters ($\boldsymbol{\beta}, \boldsymbol{\gamma}'$, REstricted Maximum Likelihood, REML, estimation). For the Poisson GLMM the integrations cannot be undertaken analytically but an approximation is available through the Laplace approximation that expands the logarithm of Eq 12 via a Taylor series to a second order, finds the values of the variables of integration at the mode of the integrand and eventually approximate the integrated joint likelihood by the value of the mode and the curvature around that point. For REML estimation the mode is found by Penalized Iterative Reweighted Least Squares, by iteratively updating the value of $\beta, \gamma'$ given the values of $\lambda, \theta$, and thus of $S_{\lambda,\theta}$, until the joint log-likelihood no longer changes according to the Newton scheme[3]:

$$\begin{aligned}
(\beta, \gamma')^{[k+1]} &= (\beta, \gamma')^{[k]} + \left((X, \Psi')^T W^{[k]}(X, \Psi') + \Sigma^{[k]}\right)^{-1} \{(X, \Psi')^T(Y - \hat{\mu}) - \Sigma^{[k]}(\beta, \gamma')^{[k]}\} \\
\Sigma^{[k]} &= diag\left(0_{dim(\beta)}, S_{\lambda,\theta}^{[k]}\right) \\
W^{[k]} &= diag(\hat{\mu}) \\
\hat{\mu} &= \exp\left(X\beta^{[k]} + \Psi'\gamma'^{[k]} + GL\right)
\end{aligned} \quad \text{Eq 14}$$

---

[2] In the case of survival frailty models, $\mathbf{Z}$ is a block diagonal matrix with rows equal to the number of (Gauss Lobatto expanded) observations and rows equal to the number of individuals. Each block is a column vector of ones which index the individuals a particular group of observations came from. The variance of $\mathbf{b}$ is equal to $\mathbf{V_\theta} = \sigma_u^2 \mathbf{I}_N$, with $\sigma_u^2$ the variance of the frailty component and $\mathbf{I}_N$ the identity matrix in $N$ dimensions.

The values of the variance components $\lambda, \theta$ may be found by numerical optimization of the REML objective:

$$\log\left(Y, \gamma'^{[k]} \middle| X, \Psi', \beta^{[k]}, GL, \lambda, \theta\right) - \left(\gamma'^{[k]}\right)^T S_{\lambda,\theta} \gamma'^{[k]}/2 - \log|S_{\lambda,\theta}^-|/2 - \log|V(X, \Psi', W^{[k]}, S_{\lambda,\theta})|/2 + \dim(\beta, \gamma') \log 2\pi/2 \quad \text{Eq 15}$$

where $V(X, \Psi', W^{[k]}, S_{\lambda,\theta}) = \begin{bmatrix} X^T W^{[k]} X & X^T W^{[k]} \Psi' \\ \Psi'^T W^{[k]} X & \Psi'^T W^{[k]} \Psi' + S_{\lambda,\theta} \end{bmatrix}$. By iterating the nested optimization scheme of Eq 14 and Eq 15, GAM software can fit the scenarios in Table 1. However, to do efficiently for frailty models, the software must be able to exploit the sparsity patterns of the matrices $\Psi'$, $\Sigma^{[k]}$, $S_{\lambda,\theta}$ and $V(X, \Psi', W^{[k]}, S_{\lambda,\theta})$. As we will see in the Results section this does not appear to be the case for the *bam* or the *gam* fitting routines. While *glmmTMB* and potentially other mixed model software are able handle sparsity by design, they lack the capability to fit models that are very large (in terms of number of observations, fixed effect parameters and smoothing components) because memory use will blow up as we attempt to accommodate large $X^T$. Applying an iterated integration scheme and the Laplace approximation to Eq 12 allows us to utilize software components that a) excel in handling large $X^T, \beta$ and $\gamma$, a task that *bam* excels at(19, 32, 35) and b) handle sparse random effect structures in $Z$ and $b$ by a modified backfitting approach. The approach is illustrated by carrying out the integration of the REML criterion corresponding to Eq 12 in stages:

$$\iiint (Y,\gamma,b|X,\Psi,Z,\beta,GL,\lambda,\theta)d\beta d\gamma db = \iiint (Y,\gamma|X,\Psi,Z,\beta,b,GL,\lambda)(b|\theta)d\beta d\gamma db$$

$$= \int \left[\iint (Y,\gamma|X,\Psi,Z,\beta,b,GL,\lambda)d\beta d\gamma\right](b|\theta)db$$

$$\iint (Y,\gamma|X,\Psi,Z,\beta,b,GL,\lambda)d\beta d\gamma \approx \frac{(Y,\hat{\gamma}|X,\Psi,Z,\hat{\beta},b,GL,\lambda)(2\pi)^{\dim(\beta,\gamma)/2}}{V(X,\Psi,\widehat{W},S_{\hat{\lambda}})}$$

$$\iiint (Y,\gamma,b|X,\Psi,Z,\beta,GL,\lambda,\theta)d\beta d\gamma db \approx (2\pi)^{\dim(\beta,\gamma)/2} \int \frac{(Y,\hat{\gamma}|X,\Psi,Z,\hat{\beta},b,GL,\hat{\lambda})}{V(X,\Psi,\widehat{W}(b),S_{\hat{\lambda}})}e^{-b^t V_\theta^{-1} b/2}db$$

$$\int \frac{(Y,\hat{\gamma}|X,\Psi,Z,\hat{\beta},b,GL,\hat{\lambda})}{V(X,\Psi,\widehat{W},S_{\hat{\lambda}})}e^{-b^t V_\theta^{-1} b/2}db \approx (2\pi)^{\dim(b)/2} \frac{(Y,\hat{\gamma}|X,\Psi,Z,\hat{\beta},\hat{b},GL,\hat{\lambda})}{V(X,\Psi,\widehat{W},S_{\hat{\lambda}})V(0,Z,\widehat{W}(\hat{b}),V_\theta^{-1})}$$

---

[3] The scheme gives the Newton update from the value of the parameters at the kth iteration to the k+1 th one.

In going from the first to the second line, we use the fact that the order of integration may be freely exchanged for the well-behaved integrands of the exponential family. In the third line, we use the fact that for fixed values of the random unit level effects $b$ (so that the entire factor $Zb + GL$, rather than the $GL$, can be treated as offset) integration over $\beta$ and $\gamma$ amounts to fitting a standard GAM, a task that can be delegated to *bam*). Having estimated $\hat{\beta}$ and $\hat{\gamma}$, the remaining integration over $b$ corresponds to a maximum likelihood estimation of a Poisson GLMM with a zero intercept, an offset equal to $X\hat{\beta} + \Psi\hat{\gamma} + GL$ and a random regression structure $Zb$. This integration can be carried out via the Laplace approximation (as we do in the fourth line) e.g. via *glmmTMB*, or in the case of individual level frailty with adaptive Gaussian quadrature, or any other appropriate integration technique (36). The following mixed software algorithm is then used to carry out the estimation as follows:

1. Expand the dataset at the Gauss Lobatto nodes and compute the Gauss Lobatto weights
2. Initialize $b$ to some value (in our experiments we initialized to zero corresponding to a survival model without frailty)
3. Compute the offsets $Zb + GL$ for the Poisson regression in the next step

4. Using *bam* optimize the REML criterion with PIRLS to obtain estimates $\hat{\lambda}, \hat{\boldsymbol{\beta}}, \hat{\boldsymbol{\gamma}}$

5. Compute the offset $X\hat{\boldsymbol{\beta}} + \Psi\hat{\boldsymbol{\gamma}} + \boldsymbol{GL}$ for the Poisson mixed model regression in the next step

6. Using *any* mixed effects software(36) that can account for the sparsity in random effects obtain estimates $\hat{\boldsymbol{b}}, \hat{\boldsymbol{\theta}}$ using Maximum Likelihood estimation as there are no fixed effects

7. Iterate Steps 1-6 until the deviance of the Poisson regression in step 6 no longer changes

To account for the bias in estimation from ignoring the remainder term of the Laplace approximation over $\hat{\lambda}, \hat{\boldsymbol{\beta}}, \hat{\boldsymbol{\gamma}}$ and speed convergence when integrating over $\boldsymbol{b}$ we applied a heuristic : instead of fitting a model without an intercept at step 6, we fit a model with a single global intercept ($\alpha$) and use REML estimation to integrate over ($\alpha, \boldsymbol{b}$).

## 2.4. Experimental design to map resource use and bias of computational methods for large scale frailty modeling in simulations

*Factors affecting method performance:* To rank methods that fit frailty models using simulations, we need to consider the factors that affect the resource use (time to deliver the result, computer memory consumption) and the bias (difference in the value of the estimate for a parameter and its actual value) when designing these simulations. For these computer experiments we used the Weibull distribution because it provides one of the simplest examples in which the GAM formulation is not exact with respect to the baseline hazard, i.e. the log-hazard is not a polynomial function. This choice allowed us to probe the robustness of the GAM approach to survival analysis with respect to the approximation error of the quadrature in Eq 9. The survival function in the Weibull model for an individual with covariate (row) vector $X_i$, frailty $z_i$ (distributed as a Gaussian random variate with mean of zero and standard deviation $\sigma_f$) and parameters $\boldsymbol{\beta}$ is given by $S(t|\lambda, \gamma, z_i, X_i, \boldsymbol{\beta}) = \exp(-\lambda \times t^\gamma \times e^{X_i\boldsymbol{\beta}} \times e^{z_i})$. In this model, the parameter $\lambda$ allows for a straightforward way to scale the event rate at any given time for a given value of the covariate vector and a fixed shape parameter $\gamma$. For this model, it appears a-priori likely that variables that can affect resource use and bias include: the number of individuals ($N$), the number of covariates, i.e. $\dim(\boldsymbol{\beta})$, the amount of frailty ($\sigma_f$), as well as the event rate and the rapidity at which events accumulate at baseline as captured by the parameters $\lambda, \gamma$. Furthermore, the type of covariates (whether binary or continue), and the distribution of their values are reasonable candidates for consideration, as they would affect the numerical condition of the design matrix, and thus the stability and convergence rate of the linear algebra algorithms used to fit the frailty models. To represent these variables in the design in a more interpretable manner, we introduce the following auxiliary variables: a) $f = 1 - Number\ of\ binary\ covariates/\dim(\boldsymbol{\beta})$, b) $S(\mathrm{T}_{max})$ : the survival at the maximum observation time in the dataset (which is fixed), c) $q = (1 - S(\mathrm{T}_{max}))/(1 - S(1))$, an index of the speed by which events accumulate in a given dataset by comparing the % of the events at the maximum observation time to that at an earlier time point and d) $r = \sqrt{Variance(X_i\boldsymbol{\beta})}/\sigma_f$, the ratio of the standard deviation of the linear predictor (computed over all individuals in a dataset with fixed $\boldsymbol{\beta}$) over that of the standard deviation of the frailty component. To motivate the parameters $r, q$ let's rewrite the Weibull survival adding and subtracting $\langle X_i\boldsymbol{\beta}\rangle$, the expectation of the linear predictor over all individuals in a dataset in the expression of the survival:

$$S(t|\lambda, \gamma, z_i, X_i, \boldsymbol{\beta}) = \exp\left(-\lambda \times t^\gamma \times e^{\langle X_i\boldsymbol{\beta}\rangle} \times e^{(X_i\boldsymbol{\beta} - \langle X_i\boldsymbol{\beta}\rangle)} \times e^{z_i}\right)$$

The quantities $(X_i\boldsymbol{\beta} - \langle X_i\boldsymbol{\beta}\rangle)$ enter the model in a manner analogous to the frailties $z_i$ : a deviation from an "average" survival given by: $\bar{S}(t|\lambda, \gamma, z_i, X_i, \boldsymbol{\beta}) = \exp\left(-\lambda \times t^\gamma \times e^{\langle X_i\boldsymbol{\beta}\rangle}\right)$. It is thus natural to introduce a dimensionless measure of the spread of the deviations $X_i\boldsymbol{\beta} - \langle X_i\boldsymbol{\beta}\rangle$ against the deviations arising from frailty, leading to the parameter $r$. On the other hand, the parameter $q$ along with $S(\mathrm{T}_{max})$ provide us with a way to think how fast the survival function goes from "top-left" $S(1)$ to its "bottom-right" value, instead of the less intuitive, but equivalent, formulation in terms of $\lambda, \gamma$.

*A D-optimal Response Surface experimental design*: In this design, we modeled the response (resource use, or bias) as a function of the variables $N, \dim(\boldsymbol{\beta}), f, r, S(T_{max}), q$ that may affect the final response in a given analysis with frailties with a Response Surface Model (RSM)(37). The parameters of the RSM were allowed to differ by the method used to estimate the frailty model and a custom D-optimal design was generated over $N, \dim(\boldsymbol{\beta}), f, r, S(T_{max}), q$ using the RSM:

$$\begin{aligned} outcome = & \quad N * dim(\boldsymbol{\beta}) * method \\ + & \quad N^2 * method + dim(\boldsymbol{\beta})^2 * method \\ + & \quad f * method + f^2 * method \\ + & \quad r * method + r^2 * method \\ + & \quad q * method + q^2 * method \\ + & \quad S(T_{max}) * method + S(T_{max})^2 * method \end{aligned}$$

where $*$ indicates the inclusion of all main effects and interaction of terms in the regression model. The RSM was generated in the hypercube $N \times \dim(\boldsymbol{\beta}) \times f \times r \times S(T_{max}) \times q = [20000, 40000] \times [10,80] \times [0.10, 0.90] \times [0.1, 1.0] \times [0.05, 0.95] \times [10, 20]$, with *method* a categorical factor with levels : *glmmTMB* as an one stage method, *bam+glmmTMB* (using *bam* for the first stage survival modeling and *glmmTMB* for the second stage) and *bam+AGQX*, with X one of 3,9,15 a two stage method that combined *bam* with Adaptive Gaussian Quadrature using an X number of nodes for the integration of the random effects (the software library *lme4*(38) was used to implement this method for all values of X). The choice of methods was motivated by preliminary experiments (detailed in the results section) that demonstrated poor resource use (execution time) for one stage frailty estimation using GAMs via the *gam* and the *bam* interfaces, as well as *coxph* and *coxme*. One thousand points for a D optimal design based on the aforementioned RSM were estimated by the R package *skpr* (39) using the code in Additional File 1. Power calculations for different standardized effect sizes (SES) are shown in Supplementary Table 1; the design was sufficiently powered at >68% to detect a moderate (SES=0.50) linear and quadratic interaction effect between method and $N$, as well as method and $\dim(\boldsymbol{\beta})$ which are expected to be the major determinants of the primary outcome considered (execution time). The secondary outcome considered for the RSM was the absolute value of the relative bias in the standard deviation of the frailty component, i.e. the absolute value of $100 \times \frac{(Estimate - True\ Value)}{True\ Value}$% that was analyzed in log scale. Analyses of the bias in the log-hazard (or more accurately log-RR) components $\boldsymbol{\beta}$ was based on a mixed model that included the RSM as a main effect, and random effects at the level of each component of $\boldsymbol{\beta}$. The design was fully replicated twice to analyze the intercept variation of the two stage models, for a total of 2,000 simulated datasets. In addition to the RSM, we also considered a simpler analysis to give an overall quantitative summary of the different methods; in this analysis, each unique combination of $N, \dim(\boldsymbol{\beta}), f, r, S(T_{max}), q$ was conceptualized as blocking factor and mixed effects models were fit to the outcome, using the block as a random effects and $method$ as the main effect. In the Supplementary Methods we discuss the simulation process to generate a dataset at each survival point and the instrumentation of the R code to collect execution time and memory usage.

## 3. Results
### 3.1. Pilot experiments for the Weibull proportional hazards model

*Poisson GAM implementations v.s. coxph* : For these experiments we simulated a clinical trial with a 1:1 assignment probability to a treatment ("trt" binary covariate), and adjusted survival estimates for three additional covariates ("adult" binary covariate present in the dataset with 60% probability and two continuous covariates "X1" and "X2"); 100 different datasets were generated for variable number of individuals (1000, 5000,10000) and event rates (controlled by the parameter $\lambda$ : 0.001 (small), 0.01 (low), 0.03 (moderate) and 0.1 (large) and different amounts of frailty, i.e. $\sigma_f$: 0 (none), 1 (low), 3 (moderate), and 7 (high). Supplementary Figure 1 illustrates the variability in survival for 100 datasets generated for different number of individuals, shape parameters and amounts of frailty under non-informative censoring (code for the figure is provided in **Additional File 2**). Compared to the scenarios without frailty, increasing amounts of frailty lead to a large proportion of individuals experiencing the event of interest fast (rapid initial descent of the Kaplan Meier curve) with a flattened appearance towards the end of follow up.

In Figure 1 (generated by the code in Additional File 3) we illustrate the relative bias of the treatment variable for various implementations of Poisson Gam fits (*bam*, *gam*, *glmmTMB*), Gauss Lobatto nodes (from 3 to 21), dataset size and event rate (values of $\lambda$) against the Cox model that corresponds to a semiparametric (SP) number of nodes in a Poisson GLM. For all models, increasing the sample size, decreased the bias irrespective of event rate. The error of all Poisson GAM models did not improve with increasing number of Gauss Lobatto nodes after the 9th. Estimates by the Poisson GAM model was indistinguishable from the gold standard *coxph* for practical purposes (differences were noted in the 3rd or fourth decimal point for ≥ 9 Gauss Lobatto nodes) as shown in Figure 2 and Supplementary Figure 2 for *glmmTMB* and *gam* implementations respectively. Estimates between the approximate, *bam*, and exact, *gam*, Poisson likelihood methods differed at the 5th decimal when the dataset was expanded with nine Gauss Lobatto nodes or more across event rates and dataset sizes (Supplementary Figure 3). Similar results were observed when *bam* estimates were contrasted to *glmmTMB* estimates (not shown).

In Figure 3 we show the execution times (time to solution in "wall-clock" seconds) for the different methods. For the Poisson GAM methodologies, increasing the number of nodes increase the time to solution with increasing number of nodes. This is the expected behavior for algorithms dominated by the time to solve iteratively re-weighted linear systems. However, there was a difference among the three methods, with *bam* being 10 times faster than either *glmmTMB* or *gam*. The performance of *coxph* also scaled nearly with the dataset size, and was 10 times faster than *bam* in the range of dataset sizes examined.

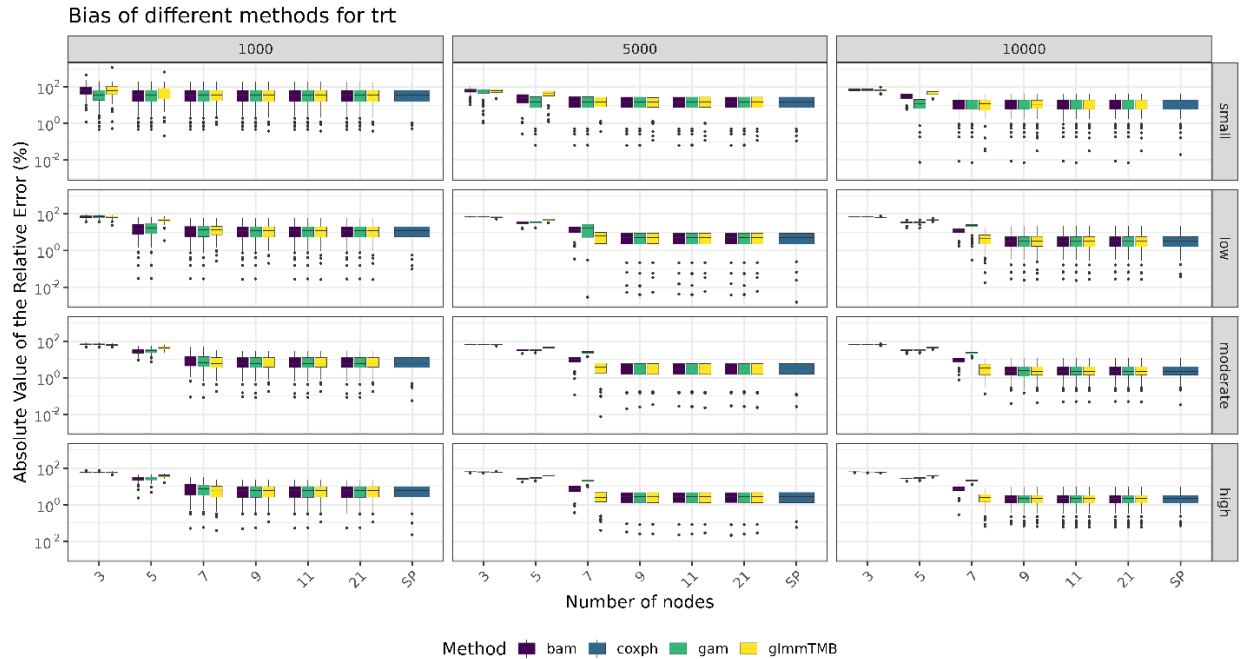

**Figure 1** Bias in estimation (absolute value of the bias as a percentage of the true value) of various implementations (*bam*, *gam*, *glmmTMB*) of Poisson GAM survival models in the absence of frailty, for an increasing number of Gauss Lobatto nodes (from 3 to 21), v.s. the Cox model (*coxph*), which corresponds to a SemiParametric (SP) Poisson fit with the number of nodes equal to the number of failures in the dataset. One hundred datasets are simulated for each combination of dataset size (1000, 5000, 10000 individuals in columns) $\lambda$: 0.001 (small), 0.01 (low), 0.03 (moderate) and 0.1 (large). The shape parameter of the Weibull was fixed at 1.2.

*In conclusion*, these pilot experiments confirm our previous observations, that the Poisson GAM method yields survival estimates that are indistinguishable from *coxph* after the dataset has been expanded with nine or more Gauss Lobatto nodes. It also demonstrates the utility of the approximate *gam* regression to cut down execution times, while yielding estimates that are not materially different compared to the exact likelihood calculations by *glmmTMB* and *gam*.

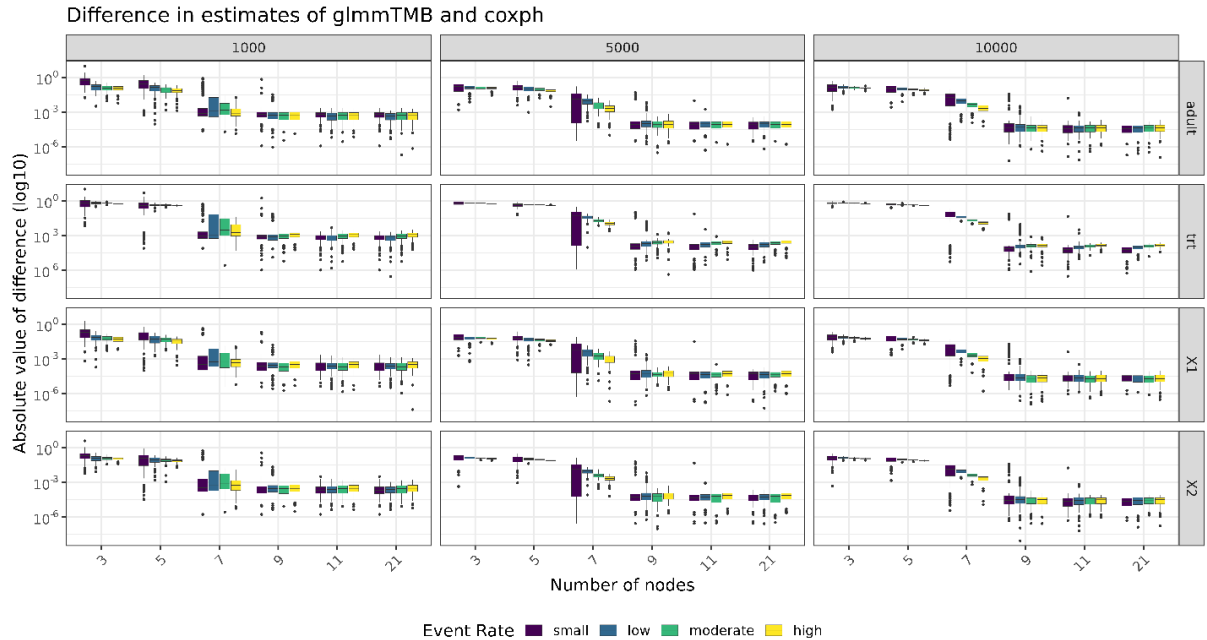

**Figure 2** Absolute difference in estimates between *glmmTMB* and *coxph* for the four covariates in the simulations from **Figure 1** and **Additional File 3**

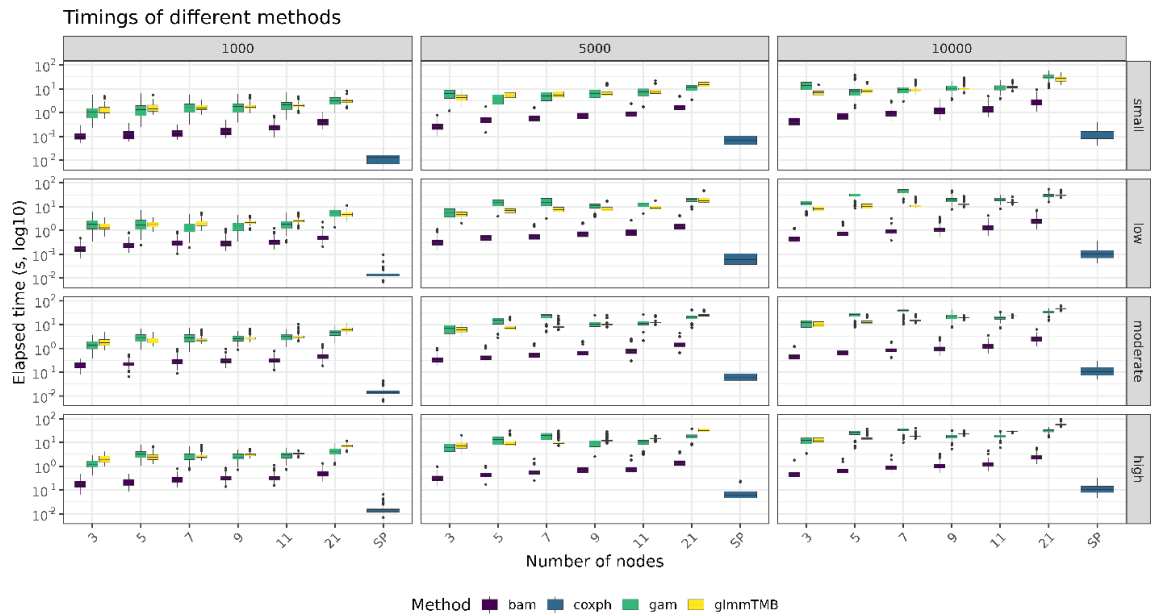

**Figure 3** Execution times for the analyses of the simulated datasets from **Figure 1** and **Additional File 3**

### 3.2. Limits of existing software for frailty modeling in moderate to large datasets

*Poisson GAM implementations without sparse estimation are non-performant.* To understand the scaling of Poisson GAM implementations for frailty models as the dataset (number of individuals and thus number of random effects) increase, we simulated 5 replicates at each combination of event rates and non-zero frailties for 1,000 and 2,000 individuals (a total of 120 datasets) using five fixed covariates as in the previous section. The raw data for these experiments in shown in Figure 4 (code for the simulation may be found in Additional File 4).

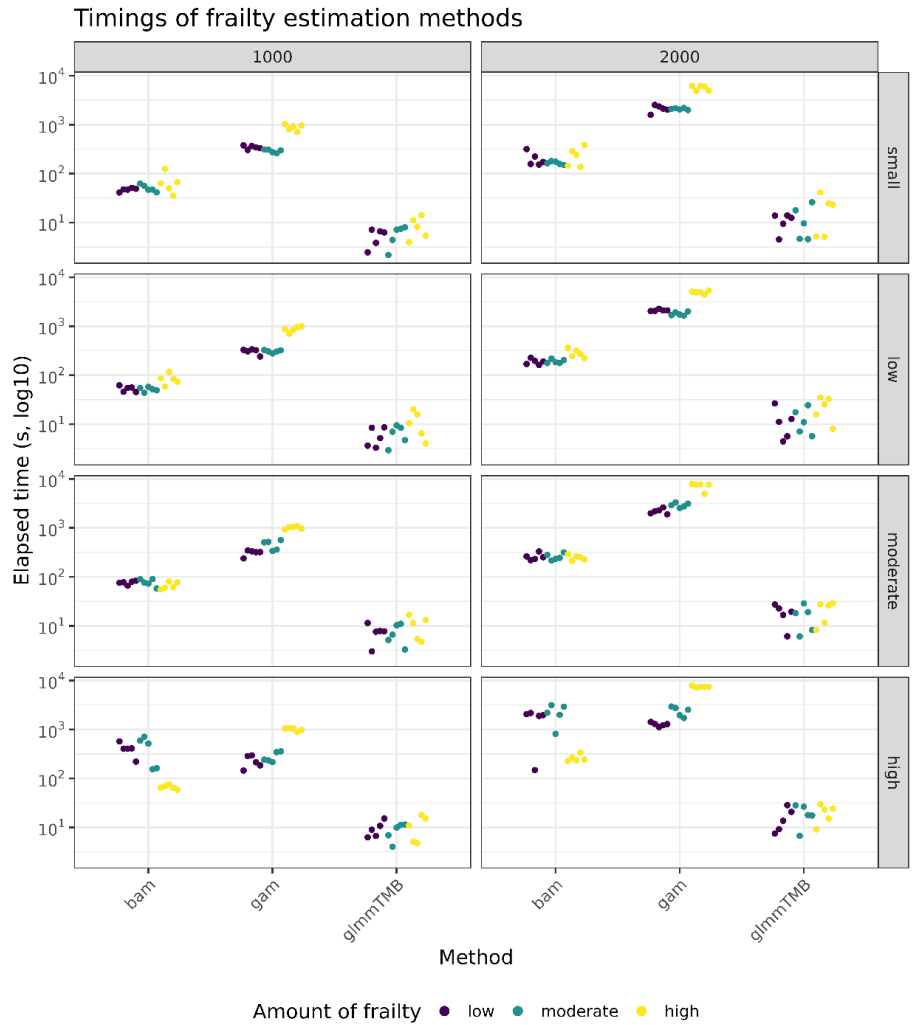

**Figure 4** Execution timing for frailty analysis of datasets with variable number of individuals (code for the simulations may be found in **Additional File 4**)

Execution timings were analyzed in log-space via a repeated measures model with method, size and of dataset and interaction of method and dataset size as fixed effects to understand the scaling of the various methods as dataset doubles. These analyses shown in Table 2 illustrate the following features: 1) using the exact method *gam*, instead of the approximate method *bam* increases execution time by over5-fold 2) *glmmTMB* reduced execution time by over 90% compared to *bam* 3) doubling the size of the dataset nearly quadrupled execution time for the *bam* and increased 12 fold (3.693 × 1.804) for *gam* 4) doubling the dataset size in *glmmTMB* resulted in doubling of execution time (3.693 × 0.535 ∼ 1.97). These findings illustrate the expected behavior, i.e. quadrupling execution time when the number of individuals is doubled, because there is double the amount of data and double the number of parameters to estimate for non-sparse estimation methods, with considerable savings of execution time when sparse estimation methods (*glmmTMB*) in this scenario. Given this scaling, and the lack of any material differences in the accuracy of the various methods (Supplementary Figure 4), we selected *glmmTMB* instead of the other methods for the moderate and big data scenarios considered later in this report. To put this decision into perspective, increasing the dataset from 1,000 to 64,000 (a 6 fold increase in log2 scale) individuals would be 4^6~4,096 slower than the baseline scenario in Table 2.

**Table 2** Execution timing for a Weibull survival model with frailty

|  | Execution Timing Estimate [95% CI] p-value |
|---|---|
| *bam* (reference, seconds)* | 83.372 [70.686, 98.333] <0.001 |
| *gam*/*bam* | 5.331 [4.221, 6.732] <0.001 |
| *glmmTMB*/*bam* | 0.086 [0.068, 0.108] <0.001 |
| x2 size in *bam* | 3.693 [2.924, 4.664] <0.001 |
| x2 size om *gam*/x2 size in *gam* | 1.804 [1.297, 2.510] <0.001 |
| x2 size in *glmmTMB*/x2 size in *bam* | 0.535 [0.384, 0.744] <0.001 |
| * reference scenario is BAM with 1000 individuals | |

*Gold standard implementations of frailty models in R (coxme, coxph) are non-performant for moderate to large datasets*

Comparisons were done between *coxph, coxme* and *glmmTMB* for datasets of 20,000 and 40,000 individuals with 20 and 40 covariates (5 replicates in a 2x2 factorial design, see Additional File 5 and

Additional File 6). The bias of the estimated frailty component was acceptable (<30%) for all methods, but it was larger for the *glmmTMB* method than either the *coxph* or *coxme* as shown in Supplementary Figure 5A. Estimates of the log-hazard components were similar across all three methods irrespective of the magnitude of the frailty component (Supplementary Figure 5B). *glmmTMB* was consistently faster in these comparisons irrespective of the amount of frailty (Figure 5A), but consumed substantially more memory (Figure 5B); in fact peak memory use by *coxph* and *coxme* was below the limit that could be reliably detected through the operating system (OS) facilities and were not considered further.

We analyzed the execution time needed to fit the three models by a linear mixed model, regressing the logarithm of the time to solution against a regression that included the logarithm of the number of individuals, the logarithm of the number of covariates and their interaction with the method as main effects, incorporating random effects for the replicate dataset. Predictions from this model are shown in Figure 5C which illustrate the impracticality of using *coxme* and *coxph* for large datasets with more than 40 covariates. Whereas *glmmTMB* is predicted to execute within one hour for a regression with forty covariates, *coxph* will require approximately five days and *coxme* nearly six weeks to provide an answer in the platform used for the simulations. The Achilles heel of *glmmTMB* in these scenarios is the predicted high memory requirement as shown in Figure 5D. A dataset with 40 covariates and 1M individuals is predicted to require nearly one Terabyte of memory. So, while *glmmTMB* 's execution time scales extremely well, its high space requirements limit the size of the datasets and number of covariates that can be fit even by high end workstations.

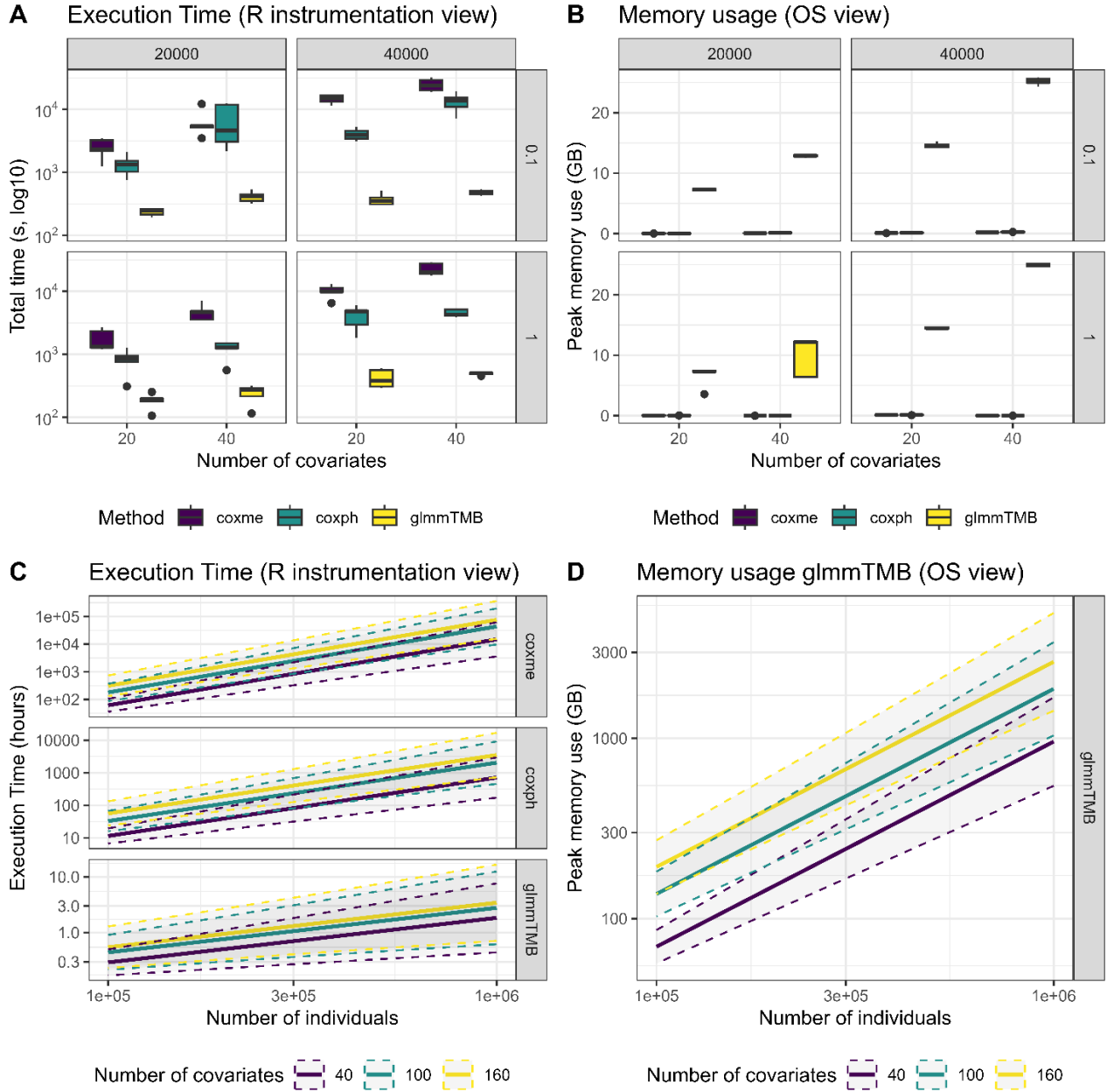

**Figure 5** Execution timings (A) and memory usage (B) for datasets of different size (20,000 and 40,000 individuals), number of covariates (20,40) and two levels of frailty (0.1 and 1 the standard deviation of the covariates) generated with the code in **Additional File 5**) and analyzed with the code in **Additional File 6**. Five replicate datasets were generated at each design point and analyzed via *coxph*, *coxme* and *glmmTMB*. The data in (A) and (B, *glmmTMB*) were analyzed via a linear (mixed) model in log space and predictions were generated for datasets of various sizes and number of covariates to assess the scalability of the three methods. Predictions of execution time and memory usage are shown in (C) and (D) respectively as estimated (sold thick lines) and 95% confidence intervals (dashed lines of the same color and light grey bands).

### 3.3. Resource surface modeling highlights the two-stage method as the performant approach for frailty modeling in moderate datasets

A regression analysis of the execution times, bias in the estimation of the variance and the log hazard components under blocking for the amount of frailty ($r$), survival at the end of the observation $S(T_{max})$, proportion of binary covariates ($f$) and ratio of events at the maximum observation time over the events at time equal to one (q) is shown in Table 3. Compared to the reference method (*glmmTMB*), the two stage methods executed faster; increasing the number of quadrature points in the adaptive integration methods resulted in decreasing bias in the estimation of the frailty variance component and the log hazard coefficients. Using an adaptive quadrature rule with 9 or 15 points resulted in a 4% and 5% smaller bias relative to the reference method.

Table 3 Execution time and estimation bias of one stage (glmmTMB) and two stage estimation methods

|  | Execution Timing | Bias in Frailty Component sd* | Bias in log-hazard Coefficients* |
| --- | --- | --- | --- |
| *glmmTMB* (reference) | 318.06 [266.80, 379.19] <0.001 | 38.08 [33.37, 43.46] <0.001 | 21.98 [19.68, 24.54] <0.001 |
| *bam + glmmTMB* (without intercept) | 0.40 [0.31, 0.50] <0.001 | 1.46 [1.24, 1.73] <0.001 | 0.87 [0.84, 0.89] <0.001 |
| *bam + AGQ03* (without intercept) | 0.38 [0.30, 0.48] <0.001 | 1.17 [1.00, 1.38] 0.056 | 1.05 [1.02, 1.08] <0.001 |
| *bam + AGQ09* (without intercept) | 0.44 [0.35, 0.56] <0.001 | 1.00 [0.84, 1.18] 0.959 | 0.97 [0.94, 0.99] 0.016 |
| *bam + AGQ15* (without intercept) | 0.41 [0.32, 0.51] <0.001 | 0.97 [0.82, 1.14] 0.688 | 0.95 [0.93, 0.98] <0.001 |
| Replicate run with intercept | 0.95 [0.77, 1.18] 0.659 | 0.98 [0.85, 1.13] 0.732 | 1.01 [0.98, 1.03] 0.645 |
| *bam + glmmTMB* (with intercept vs without intercept) | 1.29 [0.95, 1.75] 0.103 | 0.79 [0.64, 0.96] 0.020 | 1.24 [1.20, 1.29] <0.001 |
| *bam + AGQ03* (with intercept vs without intercept) | 1.35 [0.99, 1.83]<br>0.055 | 1.21 [0.99, 1.49]<br>0.063 | 1.12 [1.09, 1.16]<br><0.001 |
| *bam + AG09* (with intercept vs without intercept) | 1.29 [0.95, 1.75]<br>0.101 | 1.26 [1.03, 1.54]<br>0.027 | 1.24 [1.20, 1.28]<br><0.001 |
| *bam + AGQ15* (with intercept vs without intercept) | 1.82 [1.34, 2.47]<br><0.001 | 1.17 [0.96, 1.44]<br>0.122 | 1.19 [1.15, 1.22]<br><0.001 |

\* Bias quantified as absolute relative bias : 100*abs(1-est/actual)
The data were generated with Additional File 7 and analyzed by the code in Additional File 8

in the estimated log-hazard components. The two-stage method that combined *bam* with *glmmTMB* achieved a smaller bias in the estimation of the log-hazard coefficients at the expense of a larger bias in the estimation of the frailty component. Inclusion of an intercept term in the second stage model, increased execution times for all stage models and inflated the bias for all AQQ methods; inclusion of the intercept in the *bam + glmmTMB* method decreased the relative bias in the estimation of the frailty component from 1.46 to 1.46 x 0.98 x 0.79 = 1.13 times that of the *glmmTMB* at the expense of a higher bias in the estimation of the log-hazard coefficients from 0.87 to 0.87 x 1.01 x 1.24 = 1.09 relative to *glmmTMB*.

In Table 4 we report the results of the Analysis of Variance based on the RSM model used to design the simulation study. The most important predictors (larger F value) were the method used (memory usage outcome), dimensionality of the covariate vector (execution time) and relative size of the frailty standard deviation (r) for the bias in the frailty components and the log-hazard coefficients. Many of the interaction terms (in particular those involving the size of the dataset or the number of covariates) by the method were also highly significant, indicating a substantial modification of the effect of a particular method by the size of the dataset (number of individuals) and the complexity of the analysis (large number of covariates). Of some interest, the number of events did appear to exert an influence on both memory usage and execution time; this can be understood by referring to any of the likelihood formulations in Section 2.1: a small number of events will simplify substantially the calculations, as the terms in the loglikelihood that are multiplied by the event indicators will drop out of the calculations because they are identically zero. For the Cox model libraries, a small number of events will yield much smaller (and fewer) risk sets, also cutting down on execution time.

Table 4 Analysis of variance tables of the response surface for resource use (memory usage), execution times, bias in the standard deviation of the frailty component and the log-hazard ratio (log – relative risk) coefficients. Results are reported as F statistic and p - value

| Factors | DF | Memory Usage (GB) | | Execution Times (sec) | | Bias in Frailty Component | | Bias in log HR Coefficients | |
|---|---|---|---|---|---|---|---|---|---|
| | | F | p | F | p | F | p | F | p |
| $N$ (main effect) | 1 | 106 | 1.2e-23 | 43 | 7.6e-11 | 3 | 8.7e-02 | 171 | 0.0e+00 |
| $\dim(\boldsymbol{\beta})$ (main effect) | 1 | 502 | 3.4e-89 | 104 | 2.6e-23 | 2 | 1.9e-01 | 3811 | 0.0e+00 |
| $method$ (main effect) | 4 | 1160 | 0.0e+00 | 19 | 4.6e-15 | 12 | 3.5e-09 | 29 | 0.0e+00 |
| $N^2$ (main effect) | 1 | 0 | 6.6e-01 | 2 | 2.0e-01 | 1 | 4.0e-01 | 83 | 0.0e+00 |
| $\dim(\boldsymbol{\beta})^2$ (main effect) | 1 | 5 | 2.7e-02 | 5 | 3.3e-02 | 0 | 5.2e-01 | 164 | 0.0e+00 |
| $f$ (main effect) | 1 | 141 | 2.5e-30 | 14 | 2.3e-04 | 1 | 2.9e-01 | 61 | 4.9e-15 |
| $r$ (main effect) | 1 | 0 | 7.5e-01 | 2 | 2.0e-01 | 652 | 6.1e-106 | 78304 | 0.0e+00 |
| $f^2$ (main effect) | 1 | 2 | 1.2e-01 | 2 | 1.8e-01 | 2 | 1.7e-01 | 18 | 2.7e-05 |
| $r^2$ (main effect) | 1 | 1 | 4.3e-01 | 2 | 1.9e-01 | 80 | 2.2e-18 | 15812 | 0.0e+00 |
| $S(T_{max})$ (main effect) | 1 | 148 | 8.7e-32 | 25 | 6.6e-07 | 19 | 1.6e-05 | 43 | 4.7e-11 |
| $q$ (main effect) | 1 | 3 | 9.7e-02 | 0 | 6.8e-01 | 1 | 4.1e-01 | 211 | 0.0e+00 |
| $S(T_{max})^2$ (main effect) | 1 | 5 | 2.5e-02 | 4 | 3.4e-02 | 6 | 1.7e-02 | 4 | 4.2e-02 |
| $q^2$ (main effect) | 1 | 0 | 5.1e-01 | 0 | 9.1e-01 | 0 | 9.3e-01 | 11 | 8.5e-04 |
| $N \times \dim(\boldsymbol{\beta})$ (interaction) | 1 | 35 | 4.2e-09 | 23 | 1.9e-06 | 1 | 4.8e-01 | 22 | 2.8e-06 |
| $N \times method$ (interaction) | 4 | 96 | 2.2e-68 | 7 | 7.8e-06 | 1 | 6.3e-01 | 10 | 9.8e-08 |
| $\dim(\boldsymbol{\beta}) \times method$ (interaction) | 4 | 448 | 5.6e-215 | 14 | 3.5e-11 | 2 | 4.7e-02 | 48 | 0.0e+00 |
| $N^2 \times method$ (interaction) | 4 | 0 | 9.7e-01 | 1 | 4.6e-01 | 1 | 3.1e-01 | 37 | 0.0e+00 |
| $\dim(\boldsymbol{\beta})^2 \times method$ (interaction) | 4 | 4 | 2.4e-03 | 1 | 5.2e-01 | 3 | 2.2e-02 | 5 | 5.8e-04 |
| $f \times method$ (interaction) | 4 | 140 | 1.0e-93 | 14 | 5.8e-11 | 1 | 3.7e-01 | 12 | 1.3e-09 |
| $r \times method$ (interaction) | 4 | 0 | 1.0e+00 | 0 | 9.7e-01 | 28 | 1.5e-21 | 50 | 0.0e+00 |
| $f^2 \times method$ (interaction) | 4 | 3 | 1.5e-02 | 2 | 6.7e-02 | 1 | 2.1e-01 | 3 | 3.0e-02 |
| $r^2 \times method$ (interaction) | 4 | 1 | 5.0e-01 | 2 | 4.5e-02 | 13 | 2.4e-10 | 15 | 3.8e-12 |
| $S(T_{max}) \times method$ (interaction) | 4 | 152 | 6.9e-100 | 13 | 5.2e-10 | 5 | 1.1e-03 | 39 | 0.0e+00 |

| | | | | | | | | |
|---|---|---|---|---|---|---|---|---|
| $q \times$ method (interaction) | 4 | 4 | 5.3e-03 | 0 | 9.7e-01 | 2 | 7.8e-02 | 5 | 2.9e-04 |
| $S(T_{max})^2 \times$ method (interaction) | 4 | 5 | 8.0e-04 | 2 | 6.0e-02 | 3 | 3.1e-02 | 2 | 1.0e-01 |
| $q^2 \times$ method (interaction) | 4 | 0 | 8.8e-01 | 0 | 1.0e+00 | 1 | 5.5e-01 | 12 | 1.3e-09 |
| $N \times \dim(\boldsymbol{\beta}) \times$ method (interaction) | 4 | 31 | 4.8e-24 | 7 | 2.0e-05 | 0 | 9.6e-01 | 4 | 4.5e-03 |

To get a better understanding of the performance of the various methods, we proceeded to fit the response surface model used to generate the design points against the execution time and bias of the estimate of the frailty components. We then generated predictions for various fixed combinations of number of individual subjects in the dataset, number of covariates and a thousand combinations of $S(T_{max}), r, q, f$, with the combinations generated as a Sobol sequence to cover uniformly the four dimensional hypercube: $[0.05, 0.95] \times [0.1, 0.3] \times [10, 20] \times [0.10, 0.90]$. These predictions are shown in Figure 6.

There are several noteworthy observations about the performance of the various methods that can be inferred from this figure. Firstly, the various methods exhibit different scaling as the dataset (numbers of individuals) and the analysis (number of covariates) grows, with the worst scaling for *glmmTMB*. The *bam+AGQ* two stage methods exhibit better scaling than the *bam + glmmTMB* two stage method except for the highest number of covariates. Secondly, the *glmmTMB* appear to exhibit higher bias in the estimation of the frailty component as the number of covariates grow. Thirdly, the high accuracy AGQ two stage methods with 9 and 15 quadrature points, as well as the two stage *bam + glmmTMB* method showed a rapid reduction in the bias as the number of covariates in the analysis increased.

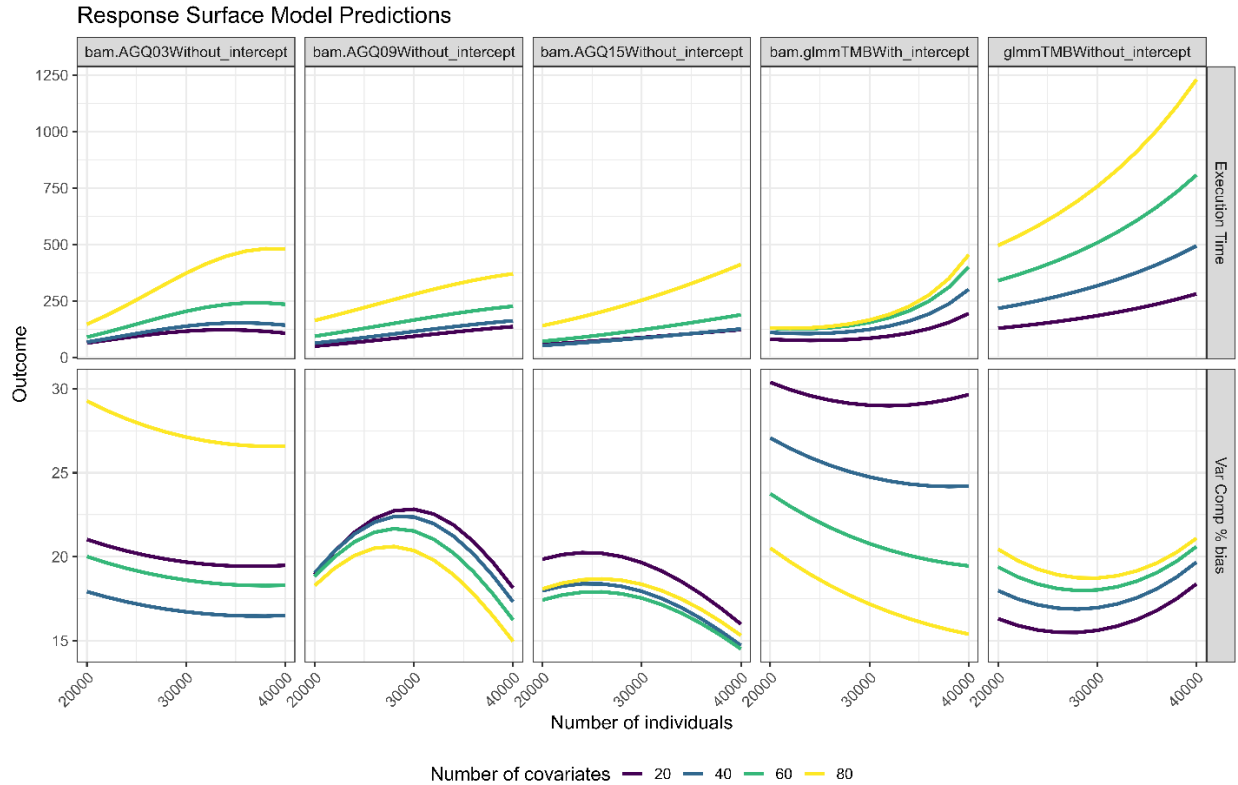

**Figure 6** Expectations of a response surface model fit to the execution time and bias in the estimate of the frailty component for a subset of the methods in **Table 3.** The figure shows the expectation for predictions at 1000 hypothetical datasets for each combination of number of individuals in the study and number of covariates. The 1000 hypothetical datasets corresponded to survival scenarios of different $S(T_{max}), r, q, f$; a Sobol sequence was used to cover uniformly the four-dimensional hypercube in $[0.05, 0.95] \times [0.1, 0.3] \times [10, 20] \times [0.10, 0.90]$

The raw memory usage data are shown in Supplementary Figure 6; similar to the analyses in Figure 5, *glmmTMB* consumed the maximum amount of memory compared to all other methods. The two stage methods exhibited memory consumption similar to *coxme* and *coxph* , i.e. they consumed small amounts, but their scaling with the size of the dataset and number of covariates were below the detection limit (~1GB) of the instrumentation code used to track memory consumption.

## 4. Concluding remarks

In this report, we extend our previous publication regarding the use of Generalized Additive Models in survival analysis and explore multiple software implementations of GAMs that allow the performant estimation of frailty survival models. The task is equivalent to that of fitting Generalized Linear Mixed Models with random effects, in which some of the random effects correspond to smoothers (e.g. the baseline hazard) and others to individual level frailties. We designed and executed numerical experiments that illustrate the pitfalls of existing software methods for frailty estimation, from either the conventional survival analysis perspective, or the GLMM/GAM one. We found that former's execution time will not scale well with the size of the dataset, or the complexity of the analysis, while the latter may fail either because of poor time (e.g. the *bam/gam* implementations) or memory (*glmmTMB*) scaling. To make these methods practical to fit for moderate sized datasets, we applied the Laplace approximation to separate the task of estimating the survival component, from that of the frailty term and this allowed us to fit datasets with very good scaling for both execution time and memory usage and acceptable bias.

While the results we obtained are very encouraging, there are a few limitations that one must keep in mind for these analyses. Firstly we did not attempt to optimize the placement of the Gauss Lobatto nodes to minimize the error of the approximation of non-polynomial hazard functions. This error probably underlines the somewhat higher bias of the methods presented herein relative to those based on the Cox model. As we showed in section 2.1, by synthesizing literature going back to the 1970s and elementary calculus, the Cox model can be given an exact likelihood interpretation for all possible baseline hazards, while the GAM approach based on Gauss Lobatto quadrature is only exact for those that are polynomials. Secondly, we didn't attempt to quantify the coverage of estimators provided by the two stage models, relying on the law of large numbers to modulate the impact of ignoring the uncertainty of the estimation of the variance components when computing the standard errors of log hazard ratios from the survival models. For the sizes of datasets (hundreds of thousands) and complexity of analyses (tens of covariates) we hope to use these methods for, the impact is likely minimal. At a practical level, there is no real alternative to the methods we propose herein for big, complex analyses, but for smaller tasks, one is probably better off using single stage methods based on survival analysis or the *glmmTMB* software. Thirdly, we didn't attempt to examine the performance of the proposed method to more complex frailty specifications, e.g. multiple failure types, shared frailty scenarios or crossed random effects. The latter are relevant in many healthcare related analyses, e.g. for example when considering subject and institutional/healthcare facility level heterogeneity. In a previous publication we have showed that TMB may offer a particularly attractive way to implement these analyses when the baseline hazard is that of the exponential distribution (26). Extensions of the work presented herein can explore the scalability of the proposed two stage methodology in these more complex random effects modeling scenarios.

## Acknowledgements

This work was supported by grants from DCI, Inc (COVID19 Vaccine Safety Study) to CA & NCATS (UL1TR001449).

# 5. Supplementary Information
## 5.1. Supplementary Methods

*Simulation of datasets of Weibull lifetimes*: Given a design point, defined by the number of individuals, the number of covariates, the proportion of binary covariates, the survival at the maximum observation time and the ratio of events at the maximum observation time over the events at time equal to one, $N, \dim(\boldsymbol{\beta}), f, r, S(T_{max}), q$, we used the following hierarchical simulation strategy to generate the datasets under non-informative censoring:

1. Calculate number of continuous $N_{cont} = \max(1, \min(\dim(\boldsymbol{\beta}) - 1, f \times \dim(\boldsymbol{\beta})))$ and binary covariates $N_{bin} = \dim(\boldsymbol{\beta}) - N_{cont}$

2. Simulate the coefficient vector corresponding to the continuous $\boldsymbol{\beta}_{cont} \sim Normal(0, \sigma_{cont}^2)$ and binary covariates $\boldsymbol{\beta}_{bin} \sim Normal(0, \sigma_{bin}^2)$. For the simulations in this report we set $\sigma_{cont} = 0.4, \sigma_{bin} = 2.0$

3. Simulate a vector of standard deviations for each of the $N_{cont}$ continuous covariate in the dataset $\boldsymbol{s}_{cont} \sim Uniform(1,3)$

4. Simulate a vector of proportions of 1 for each of the $N_{bin}$ binary covariates $\boldsymbol{p}_{bin} \sim Uniform(0.1, 0.9)$

5. For $i = 1, \ldots, N_{cont}$, simulate the ith column of the design matrix as $\boldsymbol{X}[1:N, i] \sim Normal(0, \boldsymbol{s}_{cont}[i])$ to get the continuous covariates of the design matrix

6. For $i = N_{cont} + 1, \ldots, \dim(\boldsymbol{\beta})$, simulate the ith column of the design matrix as $\boldsymbol{X}[1:N, i] \sim Bernoulli(\boldsymbol{p}_{bin}[i - N_{cont}])$ to get the binary covariates of the design matrix

7. Form the $N$ dimensional linear predictor vector $\boldsymbol{X} \begin{bmatrix} \boldsymbol{\beta}_{cont} \\ \boldsymbol{\beta}_{bin} \end{bmatrix}$ and treating the element as a random sample, calculate the sample variance $u^2 = Var\left(\boldsymbol{X} \begin{bmatrix} \boldsymbol{\beta}_{cont} \\ \boldsymbol{\beta}_{bin} \end{bmatrix}\right)$ and the sample expectation $B = E\left(\boldsymbol{X} \begin{bmatrix} \boldsymbol{\beta}_{cont} \\ \boldsymbol{\beta}_{bin} \end{bmatrix}\right)$

8. Sample the vector of $N$ frailty components as $\mathbf{z} \sim Normal(0, r^2 u^2)$

9. Compute the expected survival at time 1, as $S(1) = 1 - (1 - S(T_{max}))/q$

10. Form a very large (e.g. 100,000 or more) Monte Carlo sample of normal random variables $\mathbf{l} \sim Normal(B, (r^2 + 1)u^2)$; for given values of $\gamma, \lambda$, the expectation of $\exp(-\lambda \times T_{max}^{\gamma} \times e^l)$ $\exp(-\lambda \times e^l)$ with respect to $l$, the elements of $\mathbf{l}$ provide Monte Carlo estimates of $\acute{S}(T_{max}|\gamma, \lambda)$ and $\acute{S}(1|\gamma, \lambda)$ respectively. Use any optimization algorithm to find the estimates of $\hat{\gamma}, \hat{\lambda}$ that minimize the sum of squares of $\left(\log \acute{S}(T_{max}|\gamma, \lambda) - \log S(T_{max})\right)^2 + \left(\log \acute{S}(1|\gamma, \lambda) - \log S(1)\right)^2$.

11. Form the total design matrix and form the total linear predictor that includes both fixed and random effects $[\mathbf{X} \quad \mathbf{z}]$ and coefficient vector $\begin{bmatrix} \boldsymbol{\beta}_{cont} \\ \boldsymbol{\beta}_{bin} \\ 1 \end{bmatrix}$ and pass them along with the estimates of $\hat{\gamma}, \hat{\lambda}$ and $T_{max}$ to the function *simsurv* of the homonymous package in R(40) that simulates Weibull lifetimes under an non-informative (administrative) censoring.

Note that the aforementioned algorithm can generate Weibull lifetimes without a frailty term, simply by setting $r=0$. A standalone implementation of the simulation algorithm is provided in Additional File 9. The following figure (which can be generated by running the code in the file, shows the generation of two datasets, one with and the other without frailty using known values of $\gamma, \lambda$ and the reconstruction of these datasets by estimating the values of $\gamma, \lambda$ from the observed values of $S(T_{max})$ and $S(1)$

*Instrumentation of R code to collect resource use and performance data*: To generate timing information, the R code for the one stage (*coxph, coxme, glmmTMB*) and the two stage methods (*bam+glmmTMB, bam+AGQ03, bam+AGQ09, bam+AGQ15*) was instrumented at the level of the R code to compute the total ("clock-wall") elapsed time needed to return an estimate of relevant parameters of the survival model. For the methods that use a Gauss – Lobatto expansion of the dataset, the time to expand the dataset was also included in the dataset. To track memory usage, additional considerations come into play: R manages memory through a garbage collected allocator and the memory tracking facilities offered by the language only track total memory allocation through the R allocator and not peak memory use. To illustrate the difference, if a task requested 10 megabytes (MB) of memory, released them back and then requested them again, R's facilities will report total usage as 20 MB but will not record the peak. To complicate matters, *glmmTMB* and other non-R foreign language libraries manage their own memory and do not request memory through R's allocator. Hence memory usage by these foreign language components is invisible to R. To address these issues, we wrote a small utility in Perl (Additional File 10 released under the MIT license) to track memory use of the R process and any external libraries used by R, by tracking the resident set size component of memory by using system calls to the operating system. Further details about this utility have been presented elsewhere, i.e. [https://chrisarg.github.io/Killing-It-with-PERL/2025/01/18/Timing-Peak-DRAM-Use-In-R-With-Perl-Part-1.html](https://chrisarg.github.io/Killing-It-with-PERL/2025/01/18/Timing-Peak-DRAM-Use-In-R-With-Perl-Part-1.html) and [https://chrisarg.github.io/Killing-It-with-PERL/2025/01/19/Timing-Peak-DRAM-Use-In-R-With-Perl-Part-2.html](https://chrisarg.github.io/Killing-It-with-PERL/2025/01/19/Timing-Peak-DRAM-Use-In-R-With-Perl-Part-2.html) . The interested reader may peruse the architecture of this utility and its interface with R in these public resources. Instrumented versions of the one and two stage methods are provided in Additional File 11.

## 5.2. Supplementary Figures

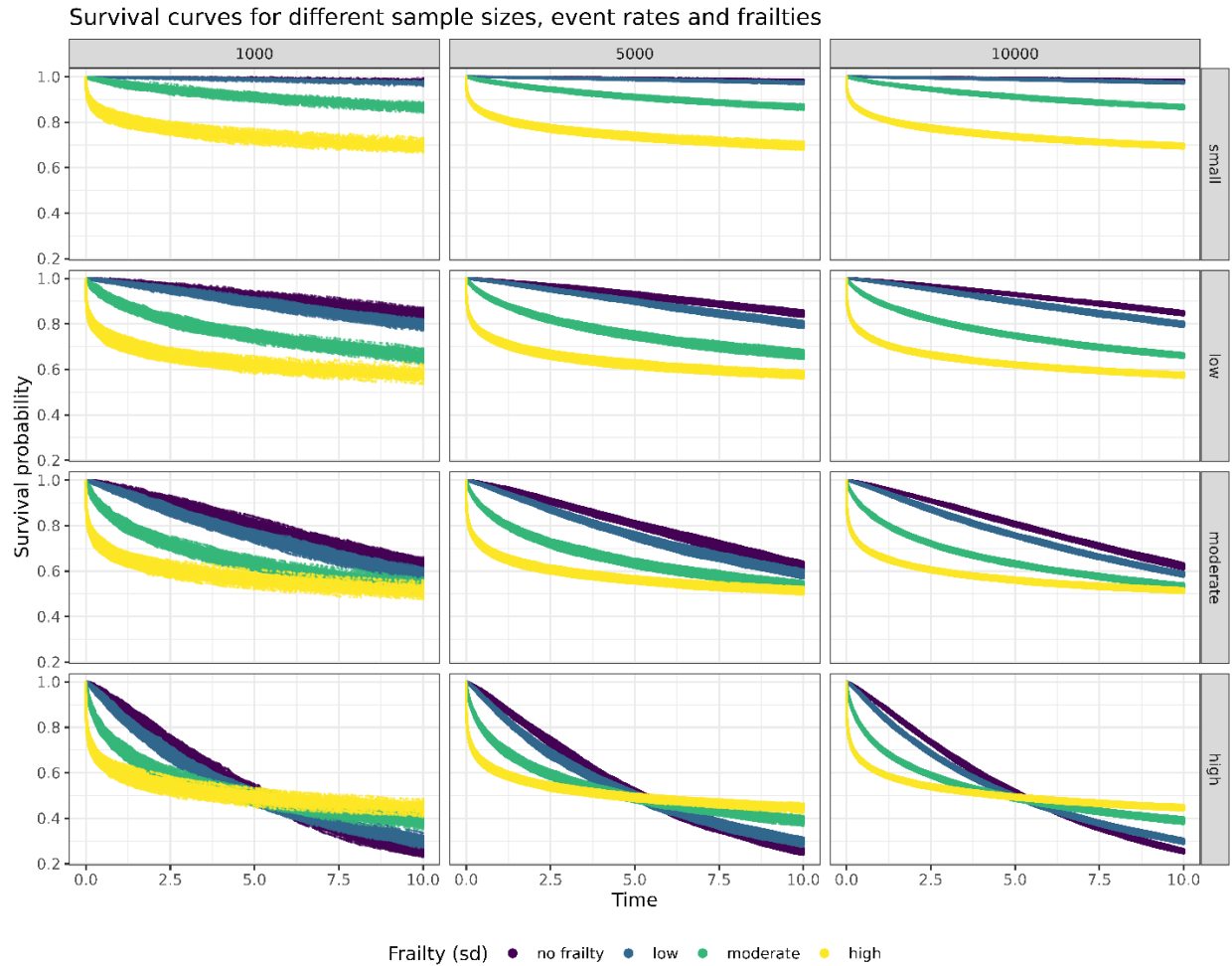

**Supplementary Figure 1** Visualizing survival for a dataset with four covariates (treatment, adult, X1, X2) for different number of participants, event rates (controlled by the parameter $\lambda$) and amounts of frailty under the Weibull proportional hazards model. One hundred datasets are simulated for each combination of dataset size (1000, 5000, 10000 individuals in columns) $\lambda$ : 0.001 (small), 0.01 (low), 0.03 (moderate) and 0.1 (large) , $\sigma_f$ : : 0 (none), 1 (low), 3 (moderate), and 7 (high). The shape parameter of the Weibull was fixed at 1.2

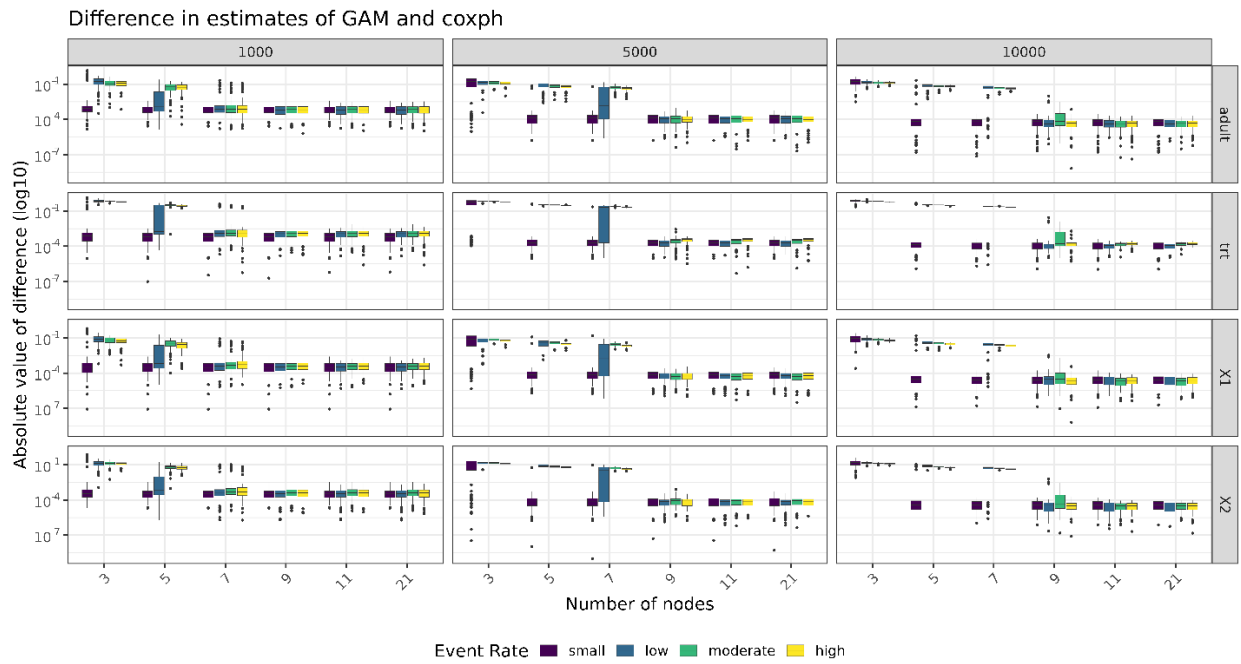

**Supplementary Figure 2** Absolute difference in estimates between *gam* and *coxph* for the four covariates in the simulations from **Figure 1** and **Additional File 3**

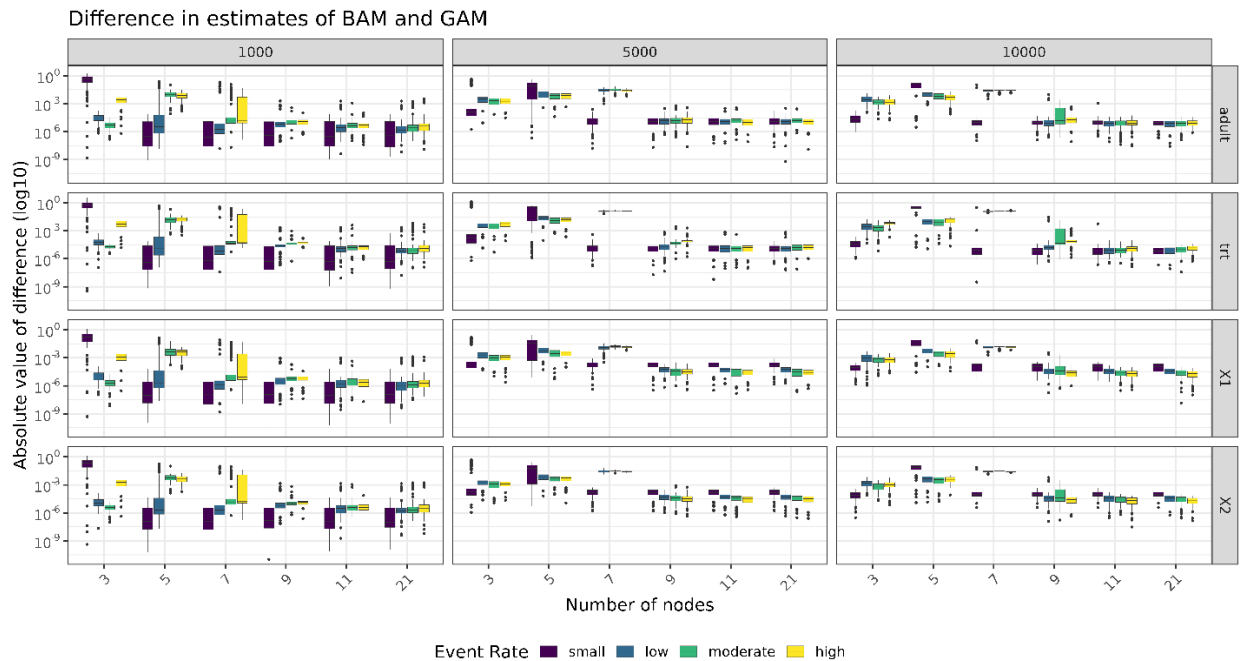

**Supplementary Figure 3** Absolute difference in estimates between *bam* and *gam* for the four covariates in the simulations from **Figure 1** and **Additional File 3**

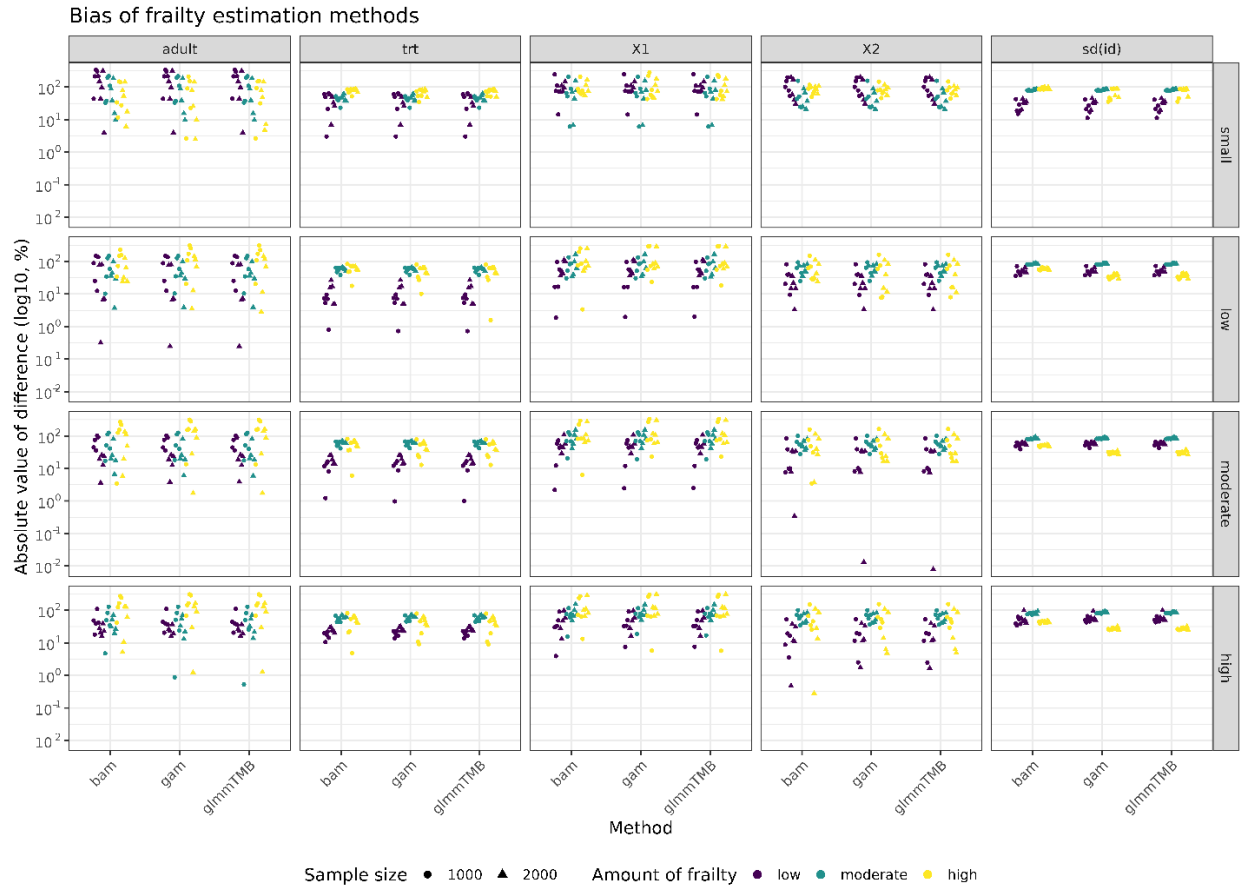

**Supplementary Figure 4** Estimates of fixed effects and standard deviation of frailty as a % of the true value for the different estimation method for Poisson GAM : *bam*, *gam*, *glmmTMB* for the simulations in **Figure 4** that was generated with **Additional File 4**

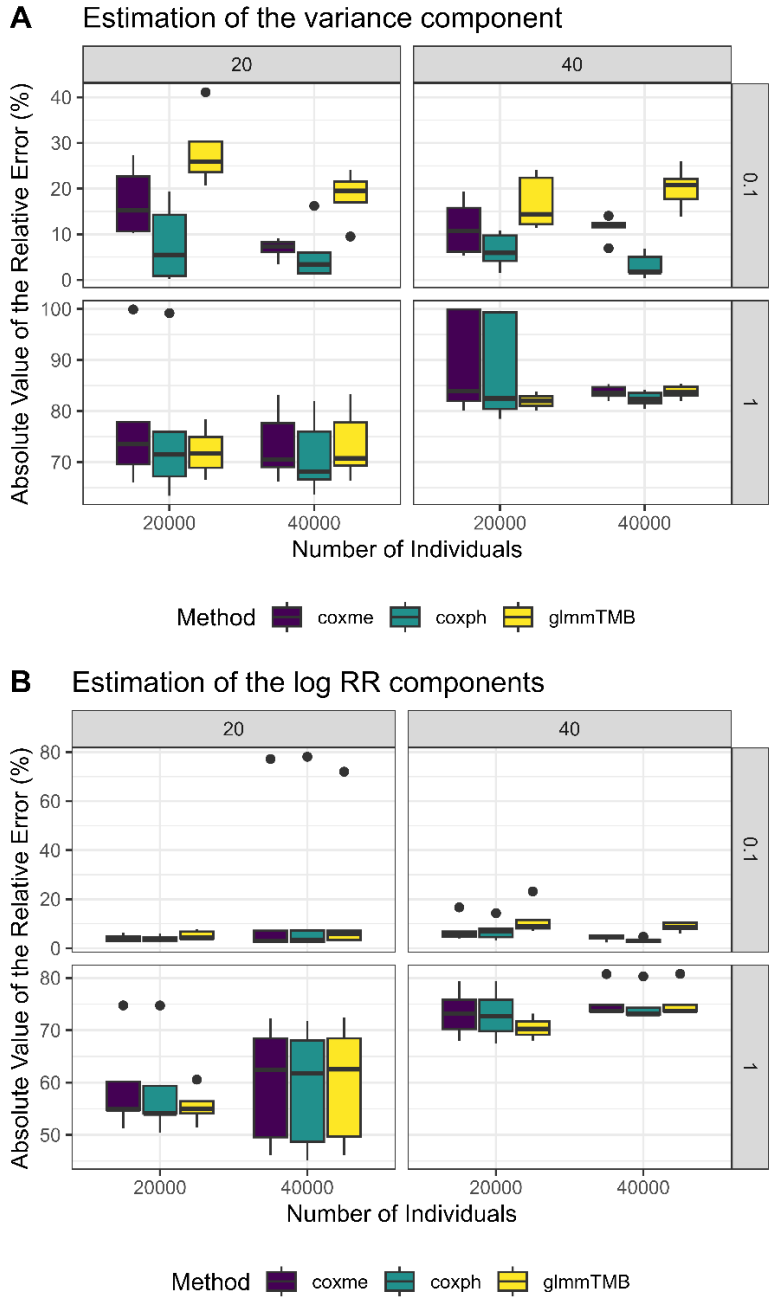

**Supplementary Figure 5** Estimates of the standard deviation of the frailty component and log-hazard ratios/log-relative risk ratios by *coxph*, *coxme* and *glmmTMB* generated with the code in **Additional File 5** and analyzed with the code in **Additional File 6** for different magnitude of the frailty standard deviation (relative to that induced by the covariates), number of individuals and number of covariates.

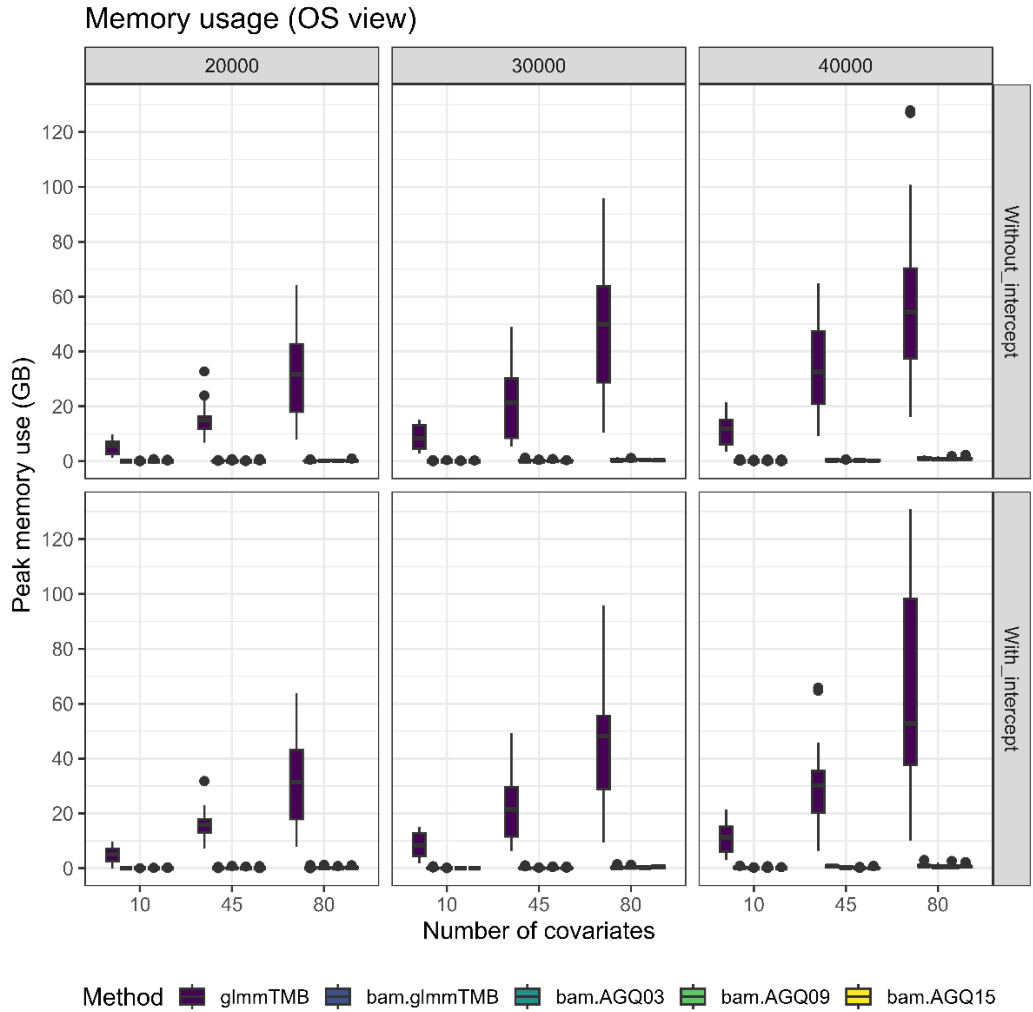

**Supplementary Figure 6** Estimates of the standard deviation of the frailty component and log-hazard ratios/log-relative risk ratios by *glmmTMB* and various two stage methods, with and without an intercept term included at the 2[nd] stage. Those were generated with the code in **Additional File 7** and analyzed with the code in **Additional File 8**.

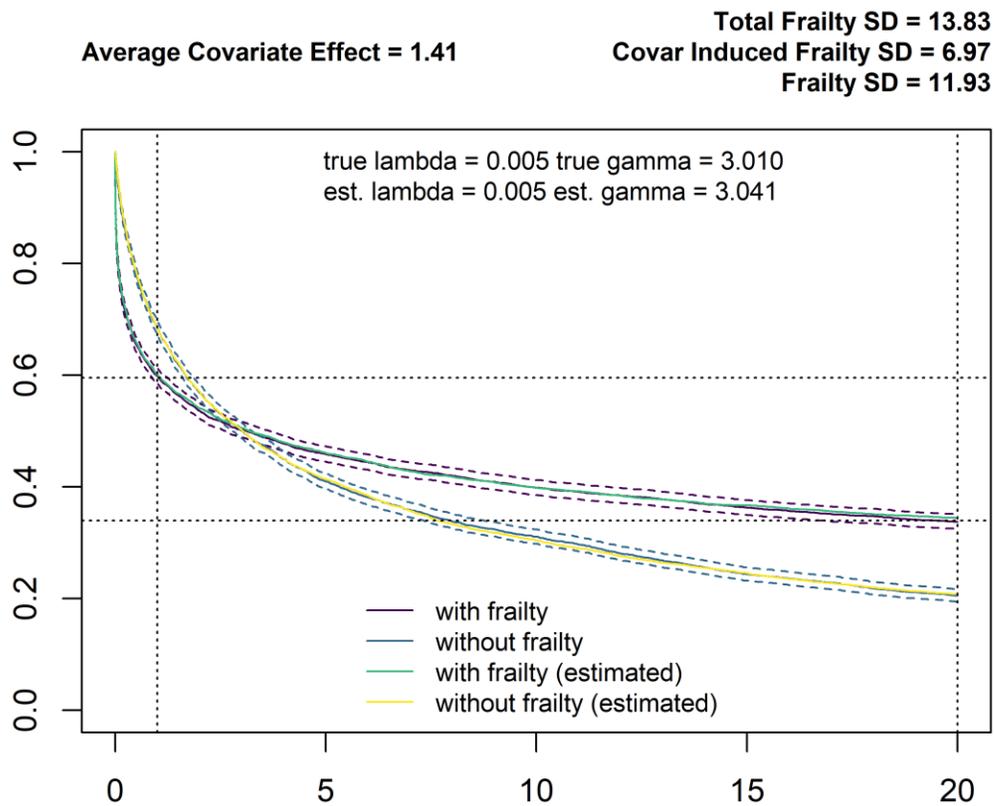

**Supplementary Figure 7** Illustration of the simulation algorithm for Weibull lifetimes using a fixed amount of frailty, under a given average covariate effect (expectation of $X\beta$ over all individuals in the dataset), covariate induced frailty standard deviation (equal to the standard deviation of of $X\beta$ over all individuals in the dataset), and a given amount of individual frailty. To generate the figure, a dataset with the given ("true") value of $\gamma, \lambda$ was generated, the survival at times 1 and 20 ($S(1), S(T_{max})$, horizontal dotted lines) was recorded and used to generate estimates $\hat{\gamma}, \hat{\lambda}$ using the algorithm in the Supplementary methods, followed by a new dataset. The figure illustrates the generation of a dataset without frailty, by setting the value of $r$, the ratio of the standard deviation of the individual frailty to that of the covariate induced induced frailty to zero. The figure was generated with **Additional File 9**.

## 5.3. Supplementary Tables

**Supplementary Table 1** Calculation for the power to detect an effect using a RSM with interactions by method to estimate frailty for different standardized effect sizes (SES)

| Effect | Power for SES = 0.25 | Power for SES = 0.50 | Power for SES = 0.75 | Power for SES = 1.0 |
|---|---|---|---|---|
| $N$ (main effect) | 0.330 | 0.859 | 0.995 | 1.000 |
| $\dim(\boldsymbol{\beta})$ (main effect) | 0.330 | 0.859 | 0.995 | 1.000 |
| $method$ (main effect) | 0.099 | 0.285 | 0.603 | 0.871 |
| $f$ (main effect) | 0.302 | 0.821 | 0.991 | 1.000 |
| $r$ (main effect) | 0.300 | 0.818 | 0.990 | 1.000 |
| $S(T_{max})$ (main effect) | 0.300 | 0.818 | 0.990 | 1.000 |
| $q$ (main effect) | 0.302 | 0.821 | 0.991 | 1.000 |
| $N^2$ (main effect) | 0.118 | 0.330 | 0.625 | 0.860 |
| $\dim(\boldsymbol{\beta})^2$ (main effect) | 0.118 | 0.330 | 0.625 | 0.860 |
| $f^2$ (main effect) | 0.132 | 0.382 | 0.701 | 0.913 |
| $r^2$ (main effect) | 0.132 | 0.384 | 0.704 | 0.915 |
| $S(T_{max})^2$ (main effect) | 0.132 | 0.384 | 0.704 | 0.915 |
| $q^2$ (main effect) | 0.132 | 0.382 | 0.701 | 0.913 |
| $N \times \dim(\boldsymbol{\beta})$ (interaction) | 0.270 | 0.767 | 0.981 | 1.000 |
| $N \times method$ (interaction) | 0.679 | 1.000 | 1.000 | 1.000 |
| $\dim(\boldsymbol{\beta}) \times method$ (interaction) | 0.678 | 1.000 | 1.000 | 1.000 |
| $N^2 \times method$ (interaction) | 0.193 | 0.675 | 0.970 | 1.000 |
| $\dim(\boldsymbol{\beta})^2 \times method$ (interaction) | 0.193 | 0.675 | 0.970 | 1.000 |
| $f \times method$ (interaction) | 0.623 | 0.999 | 1.000 | 1.000 |
| $r \times method$ (interaction) | 0.624 | 0.999 | 1.000 | 1.000 |
| $f^2 \times method$ (interaction) | 0.227 | 0.765 | 0.989 | 1.000 |
| $r^2 \times method$ (interaction) | 0.226 | 0.764 | 0.989 | 1.000 |
| $S(T_{max}) \times method$ (interaction) | 0.624 | 0.999 | 1.000 | 1.000 |
| $q \times method$ (interaction) | 0.625 | 0.999 | 1.000 | 1.000 |
| $S(T_{max})^2 \times method$ (interaction) | 0.226 | 0.764 | 0.989 | 1.000 |
| $q^2 \times method$ (interaction) | 0.226 | 0.763 | 0.989 | 1.000 |
| $N \times \dim(\boldsymbol{\beta}) \times method$ (interaction) | 0.559 | 0.996 | 1.000 | 1.000 |

## 5.4. Supplementary R code files

**Additional File 1** ("design_comparison_study"): Generate the design for the Response Surface Model used to analyze resource use and performance of one and two stage frailty models.

**Additional File 2** ("simulate_visualize_survival_datasets.R"): Code that simulates and visualizes the survival for a dataset with four covariates (treatment, adult, X1, X2) for different number of participants, event rates (controlled by the parameter $\lambda$) and amounts of frailty under the Weibull proportional hazards model.

**Additional File 3** ("simulate_survival.R"): Simulate variable number of Weibull variates in pilot experiments of a randomized controlled trial adjusting for four covariates (treatment, adult, X1, X2) for different number of participants and event rates (controlled by the parameter $\lambda$) under no frailty and analyze them via the Poisson GAM approach (functions *bam/gam* in package *mgcv*, *glmmTMB* in the homonymous package) and the Cox proportional hazards model (function *coxph* in the *survival* package)

**Additional File 4** ("simulate_timing_glmmTMB_BAM_GAM_RE.R"): Simulate datasets for a randomized controlled trial adjusting for four covariates (treatment, adult, X1, X2) for different number of participants and event rates (controlled by the parameter $\lambda$) under variable amount of frailty and analyze them via the Poisson GAM approach (functions *bam/gam* in package *mgcv*, *glmmTMB* in the homonymous package

**Additional File 5** ("compare_coxph_coxme_glmmTMB.R"): Simulate datasets of 20,000 and 40,000 individuals with 20 and 40 covariates under a 2 x 2 factorial design. Five replicates for each design point were analyzed for two different values of frailty and the execution timing of *coxme*, *coxph* (with gaussian frailty) and *glmmTMB* were compared.

**Additional File 6** ("analyze_coxph_coxme_glmmTMB.R"): Plot results of experiments in simulated datasets of 20,000 and 40,000 with 20 and 40 covariates under a 2 x 2 factorial design using **Additional File 5**. Five replicates for each design point were analyzed for two different values of frailty and the execution timing of *coxme*, *coxph* (with gaussian frailty) and *glmmTMB* were compared.

**Additional File 7** ("simulate_frailty.R"): Simulate datasets for the response surface analyses. A total of 1000 datasets were generated for the analyses with and without dataset at the design points generated by the **Additional File 1**, using the simulation facilities provided by the **Additional File 11. The** datasets were then fit to the one and two stage models using the instrumented code provided by the fitting routines in **Additional File 11**

**Additional File 8** ("analyze_response_intercept.R"): Analyzed the response surface measurements generated by **Additional File 7**.

**Additional File 9** ("explore_simulation_strategy.R"): Illustrates the simulation algorithm for Weibull lifetimes

**Additional File 10** ("monitor_memory.pl"): Perl utility to monitor memory of the parent process that calls this utility.

**Additional File 11** ("simulate_frailty_methods.R"): Simulates Weibull datasets and provides the fully instrumented one stage (*coxme, coxph, glmmTMB*) and two stage (*bam+glmmTB*, *bam+AGQX,* X 3,9, 15) methods for the estimation of frailty components. It is intended to be used by **Additional File 7**.